\documentclass[3p,times,10pt]{elsarticle}
\usepackage{amsmath}
\usepackage{amssymb}
\usepackage{hyperref}
\hypersetup{colorlinks,
	citecolor=black,
	filecolor=black,
	linkcolor=black,
	urlcolor=black
}

\usepackage{float}
\usepackage{multirow}
\usepackage{tikz}
\usepackage{caption}
\usepackage{subcaption}
\usepackage{algorithm}
\usepackage{algpseudocode}
\usepackage{listings}
\usepackage{xcolor}
\usepackage{xspace}
\usepackage{pgfplots}
\usepackage{cleveref}
\usepackage{bm}
\usepackage[symbol]{footmisc}

\usepackage[figuresright]{rotating}
\usepackage{ecrc}
\usetikzlibrary{external}
\usepackage{soul}
\usepackage[textsize=tiny]{todonotes}

\makeatletter
\renewcommand{\todo}[2][]{\tikzexternaldisable\@todo[#1]{#2}\tikzexternalenable}
\makeatother

\newcommand{\norm}[1]{\left\lVert#1\right\rVert}

\newcommand{\myint}{\int\limits}
\newcommand{\diff}[1]{\, d#1}
\newcommand{\vect}[1]{\boldsymbol{#1}}
\usepackage{mleftright}
\newcommand{\of}[1]{\mleft( #1 \mright)}

\newcommand{\reals}{\mathbb{R}}

\newcommand{\vr}{v}

\newcommand{\vtheta}{{v_{\theta}}}
\newcommand{\vphi}{v_{\varphi}}
\newcommand{\utheta}{{u_{\theta}}}
\newcommand{\uphi}{u_{\varphi}}

\newcommand{\vrunit}{\hat{\vect{v}}_{r}}
\newcommand{\vthetaunit}{\hat{\vect{v}}_{\theta}}
\newcommand{\vphiunit}{\hat{\vect{v}}_{\varphi}}

\newcommand{\bolsig}{\textsc{Bolsig+}\xspace}

\definecolor{codegreen}{rgb}{0,0.6,0}
\definecolor{codegray}{rgb}{0.5,0.5,0.5}
\definecolor{codepurple}{rgb}{0.58,0,0.82}
\definecolor{backcolour}{rgb}{0.95,0.95,0.92}

\lstdefinestyle{mystyle}{
  backgroundcolor=\color{backcolour}, commentstyle=\color{codegreen},
  keywordstyle=\color{magenta},
  numberstyle=\tiny\color{codegray},
  stringstyle=\color{codepurple},
  basicstyle=\ttfamily\footnotesize,
  breakatwhitespace=false,         
  breaklines=true,                 
  captionpos=b,                    
  keepspaces=true,                 
  numbers=left,                    
  numbersep=5pt,                  
  showspaces=false,                
  showstringspaces=false,
  showtabs=false,                  
  tabsize=2
}

\definecolor{a1}{RGB}{228,26,28}
\definecolor{a2}{RGB}{55,126,184}
\definecolor{a3}{RGB}{77,175,74}
\definecolor{a4}{RGB}{152,78,163}
\definecolor{a5}{RGB}{255,127,0}
\definecolor{a6}{RGB}{166,86,40}
\definecolor{a7}{RGB}{166,86,40}
\definecolor{a8}{RGB}{247,129,191}

\makeatletter
\def\ps@pprintTitle{%
	\let\@oddhead\@empty
	\let\@evenhead\@empty
	\let\@oddfoot\@empty
	\let\@evenfoot\@oddfoot
}
\makeatother

\volume{00}
\firstpage{1}
\runauth{}

\begin{document}

\begin{frontmatter}



\dochead{}

\title{A fast solver for the spatially homogeneous electron Boltzmann equation \footnote[2]{This work was funded by the U.S. Department of Energy, National Nuclear Security Administration award number DE-NA0003969.}}
\author{Milinda Fernando\corref{a1}}
\ead{milinda@oden.utexas.edu}

\author{Daniil Bochkov\corref{a1}}
\ead{bockkov.ds@gmail.com}

\author{James Almgren-Bell\corref{a1}}
\ead{jalmgrenbell@utexas.edu}

\author{Todd Oliver\corref{a1}}
\ead{oliver@oden.utexas.edu}

\author{Robert Moser\corref{a1}}
\ead{rmoser@oden.utexas.edu}

\author{Philip Varghese\corref{a1}}
\ead{varghese@mail.utexas.edu}

\author{Laxminarayan Raja\corref{a1}}
\ead{lraja@mail.utexas.edu}

\author{George Biros\corref{a1}}
\ead{biros@oden.utexas.edu}

\cortext[a1]{The University of Texas at Austin}




\begin{abstract}

We present a numerical method for the velocity-space, spatially homogeneous, collisional Boltzmann equation for electron transport in low-temperature plasma (LTP) conditions. Modeling LTP plasmas is useful in many applications, including advanced manufacturing, material processing, semiconductor processing, and hypersonics, to name a few. Most state-of-the-art methods for electron kinetics are based on Monte-Carlo sampling for collisions combined with Lagrangian particle-in-cell methods. We discuss an Eulerian solver that approximates the electron velocity distribution function using  spherical harmonics (angular components) and B-splines (energy component). Our solver supports electron-heavy elastic and inelastic binary collisions, electron-electron Coulomb interactions, steady-state and transient dynamics, and an arbitrary nmber of angular terms in the electron distribution function. We report convergence results and compare our solver to two other codes: an in-house particle Monte-Carlo method;  and \bolsig, a state-of-the-art Eulerian solver for electron transport in LTPs. Furthermore, we use our solver to study the relaxation time scales of the higher-order anisotropic correction terms. Our code is open-source and provides an interface that allows coupling to multiphysics simulations of low-temperature plasmas. 

\end{abstract}


\end{frontmatter}



\section{Introduction} \label{sec:intro}
Low-temperature plasma (LTP) simulations are used in many science and engineering applications. Examples include understanding plasma-surface interactions for advanced manufacturing and semiconductor processing, modeling atomic layer processing, designing medical devices, optimizing plasma-assisted combustion and chemical conversion, and simulating ionized gases in aerospace engineering~\cite{Adamovich_2017}. Simulation of low-temperature weakly-ionized        plasmas is a challenging task due to non-equilibrium chemistry, multiple time scales, and non-local electromagnetic coupling of charged particles. Electrons in LTPs (i.e., gas discharge plasmas) often deviate from thermal equilibrium, causing their velocity space distribution to deviate from the Maxwell-Boltzmann distribution. Accurate kinetic treatment of electrons is crucial because electron-induced reaction rates are highly sensitive to their tail distribution. Representing an arbitrary electron (velocity) distribution function (EDF) necessitates solving the Boltzmann transport equation (BTE) for electrons.

We consider the electron Boltzmann transport equation with an imposed electric field $\vect{E}\of{\vect{x}, t}$ given by~\Cref{eq:bte-full}. \Cref{eq:bte-full} serves as the governing equation for EDF $f(\vect{x},\vect{v},t)$, where $\vect{x}\in \Omega \subset \reals^3$ is position,  $\vect{v}\in \reals^3$ is velocity, and $t\in \reals$ is time.
\begin{equation}\label{eq:bte-full}
    \partial_t f + \vect{v} \cdot \nabla_{\vect{x}} f- \frac{q_e \vect{E}}{m_e} \cdot \nabla_{\vect{v}} f = C_{en}(f) + C_{ee}(f,f).
\end{equation}

In kinetic simulations of low-temperature plasmas~\cite{cercignani1988,hagelaar2005solving}, one seeks to solve the BTE for the EDF $f\of{\vect{x}, \vect{v}, t}$. In \Cref{eq:bte-full}, $q_e$ is the electron charge, $m_e$ is the electron mass, $C_{en}$ is a linear integral operator on $f$ representing all electron-heavy binary collisions (e.g., elastic, excitation, ionization), and $C_{ee}$ is a nonlinear operator representing the electron-electron Coulomb interactions. We restrict our attention to the spatially homogeneous case in which $f$ is only a function of $\vect{v}$ and $t$ (i.e., $\nabla_{\vect{x}} f = 0$), and $\vect{E}\equiv \vect{E}\of{t}$ is only a function in time. The homogeneous BTE for $f=f(\vect{v},t)$ is then given by \Cref*{eq:bte} and solved for $\vect{v} \in \mathbb{R}^3$ given an initial condition $f(\vect{v},0)$.
\begin{equation}\label{eq:bte}
    \partial_t f - \frac{q_e \vect{E}}{m_e} \cdot \nabla_{\vect{v}} f = C_{en}(f) + C_{ee}(f,f).
\end{equation}
In low-temperature plasmas, \Cref{eq:bte} describes electron kinetics (i.e., collision rates, mobility and diffusivity) and is coupled to a continuum or, as it is commonly referred to in the LTP literature, a \emph{``fluid''} partial differential equation (PDE) for the evolution of the \emph{``heavies''}: ground states, ions, and excited/metastable states~\cite{panneer-raja2015}. Most LTP solvers are based on operator split schemes which involve a step with a standalone solve for \Cref{eq:bte} for every spatial point $\vect{x}_i$. Solving \Cref{eq:bte} is challenging due to the combination of advection and collision terms, the multiple time scales, and the nonlinearity due to the $C_{ee}$ term. Here we focus on this BTE solve.

There are two main numerical approaches to solving \Cref{eq:bte}. The first approach is the use of the particle-based, direct simulation Monte Carlo (DSMC) method. In the above, underlying collisions are approximated with Monte Carlo sampling, and Lagrangian advection~\cite{caflisch1998monte} is used for particle advection in phase space.  
The second approach is to directly discretize \Cref{eq:bte}, typically using an Eulerian scheme for the advection~\cite{dimarco2014numerical}.
Since we are not using sampling, we also refer to such an approach as a deterministic solver. There are additional approaches that are hybrid methods~\cite{mieussens00,morris2011monte}.  All these methods have advantages and disadvantages~\cite{dimarco2014numerical}. For LTPs, some discussion can be found in~\cite{alves2018foundations}.

Most Eulerian LTP solvers use spherical coordinates $\vect{v} = \{\vr,\vtheta,\vphi\}$ where $v_r$, $\vtheta$, and $\vphi$ denote radial, polar, and azimuthal velocity space coordinates respectively. Further, in a low-temperature regime, they assume azimuthal symmetry (i.e., under additional assumptions) ($f(\cdot,\cdot,\vphi)=$ constant) and use a two-term expansion of $f$ in $\vtheta$, which reads $f(\vect{v},t) = f(\vr,\vtheta,t) = f_0(\vr,t) + f_1(\vr,t)\cos(\vtheta)$. Even with these simplifications, computing numerical solutions for the Boltzmann equation can be challenging due to the complexity of underlying collisions and the need for the accurate representation of the tails of the EDF.

Our solver follows the Eulerian approach and a Galerkin discretization scheme. We use B-splines in the radial direction and spherical harmonics in the angular directions. The main contributions of this work are summarized below.

\begin{itemize}
    \item \textbf{Multi-term azimuthal approximation}: Our solver supports arbitrary terms in both polar and azimuthal directions and thus enables the efficient and accurate representation of highly anisotropic EDFs.  (see~\Cref{sec:pde})
    \item \textbf{Steady state \& transient solutions}: We introduce a constrained Newton scheme for computing steady-state solutions and an implicit time stepping for  transient solutions. (see~\Cref{sec:pde})
    \item \textbf{Support for Coulomb interactions}: The effect of the electron-electron Coulomb interactions is significant in low-temperature plasma simulations with significant degree of  ionization~\cite{hagelaar2015coulomb}. We support electron-electron interactions with multi-term expansion and a symbolic code generation framework to autogenerate and sparsify discretized Coulomb tensors. (see~\Cref{sec:pde})
    \item \textbf{Cross-code verification}: We present detailed accuracy and verification results with existing state-of-the-art \bolsig~and an in-house DSMC code. (see~\Cref{sec:dsmc} and \Cref{sec:results})
    \item \textbf{Open-source}: Our implementation is open-source, and available in \href{https://github.com/ut-padas/boltzmann}{https://github.com/ut-padas/boltzmann}.
\end{itemize}

\textbf{Related work}: There exist several codes for solving the spatially homogeneous Boltzmann equation with the two-term approximation~\cite{hagelaar2005solving, frost1962rotational, COMSOL1998, pancheshnyi2008zdplaskin}. Eulerian solvers include \textsc{ELENDIF}~\cite{morgan1990elendif} uses two-term approximation $f(v,\vtheta) = f_0(v) + f_1(v)\cos(\vtheta)$ for the electron distribution function with capabilities for inelastic, electron-electron and several other collisional processes. The well-known \bolsig~\cite{hagelaar2005solving,hagelaar2015coulomb} uses the same two-term approximation with exponential finite differences. \text{ZDPlasKin}~\cite{pancheshnyi2008zdplaskin} is another tool for plasma kinetic simulations where \bolsig~code is used as the underlying Boltzmann solver.
\cite{mutibolt} is another work that supports multiterm approximations. However, it does not support Coulomb interactions and transient solutions. 
There have been many works~\cite{gamba2018galerkin, alekseenko2014deterministic} on using the Galerkin approach for solving the Boltzmann equation. The above works mainly rely on the analytical cross-section approximations for underlying collisions, with no support for Coulomb interactions, ionization, and other complex collisions. In~\cite{gamba2018galerkin} the authors propose global polynomial approximations for the energy coordinate direction. However, global approximations do not work for non-smooth cross-sections, such as the excitation and ionization cross-sections.

Direct simulation Monte Carlo (DSMC)~\cite{bird1994molecular} is another state-of-the-art approach for solving the Boltzmann equation. There exist several DSMC codes~\cite{scanlon2010open, zabelok2015adaptive, frezzotti2011solving, oblapenko2020velocity, bartel2003modelling, lymberopoulos1994stochastic, levko2021vizgrain} in the literature. In this work, we use our own DSMC code, details of which are provided in \Cref{sec:dsmc}~\cite{vahedi1995monte,PhysRevE.55.4642}. In addition to our algorithmic contributions, our deterministic solver is available as an open source. To our knowledge, this is the only available multi-term Eulerian electron BTE solver for LTPs.  Being open source, our solver can be easily coupled to LTP fluid approximations, as opposed to the usual tabulation of results, which do not scale well with the number of excited species. Furthermore, our code allows transient simulations.

\section{Methodology} \label{sec:methodology}

\subsection{Boltzmann equation}
\label{subsec:bte}
The Boltzmann equation is a nonlinear integro-differential equation that describes the evolution of the electron velocity distribution function $f$. At time $t$, $f(\vect{x}, \vect{v}, t) d\vect{x} d\vect{v}$ represents the number of electrons in the phase-space volume element $d\vect{x} d\vect{v}$. As mentioned, we limit our focus to spatially homogeneous distribution functions (i.e., $\nabla_{\vect{x}} f = 0$). Under the above conditions, the governing equation for the electron distribution function $f$ is given by \eqref{eq:bte}. The left-hand-side (LHS) of \eqref{eq:bte} denotes the acceleration of the electrons due to the electric field, and the right-hand-side (RHS) operators $C_{en}$ and $C_{ee}$ denotes the underlying electron-heavy collisions and electron-electron Coulomb interactions respectively.  

	\subsection{Eulerian solver}\label{sec:pde}
	We use a Galerkin discretization for \Cref{eq:bte}. We use spherical coordinates, with compactly supported B-splines in radial coordinates with real-valued spherical harmonics in the angular directions to represent EDF $f$ as specified by \Cref{eq:f_expansion} where the spherical harmonics are defined in \Cref{eq:sph_harmonics}. We use uniformly spaced knots in speed to represent a sequence of B-spline basis functions for the radial coordinate.

\begin{equation}
    f(\vect{v}, t) = \sum_{klm} f_{klm}\of{t} \Phi_{klm}\of{\vect{v}}, \text{ where } \Phi_{klm}\of{\vect{v}}  = \underbrace{\phi_k\of{v}}_{\text{B-Splines}} \overbrace{Y_{lm}\of{v_\theta, v_\phi}}^{\tiny\text{spherical harmonics}}; \label{eq:f_expansion} 
\end{equation}
and
\begin{equation}
    Y_{lm}\of{\vtheta,\vphi} = U_{lm} P^{|m|}_l\of{\cos\of{\vtheta}} \alpha_m\of{\vphi} \text{ , } U_{lm} = 
    \begin{cases}
        (-1)^m \sqrt{2} \sqrt{\frac{2l+1}{4\pi} \frac{(l-|m|)!}{(l+|m|)!}}, &m\neq 0, \\
        \sqrt{\frac{2l+1}{4\pi}}, &m = 0,
    \end{cases} \text{ , }
    \alpha_m\of{\vphi}  =
        \begin{cases}
            \sin\of{|m|\phi}, &m < 0, \\
            1, &m = 0,\\
            \cos\of{m\phi}, &m > 0.
        \end{cases} \label{eq:sph_harmonics}
\end{equation}
We select the velocity space $z$-axis is aligned with the given $\vect{E}(t)$ field. The electric field triggers polar-angle anisotropy for $f$ while maintaining azimuthal symmetry. The azimuthal symmetry is naturally represented in spherical harmonics by setting $m=0$ and varying $l$ as desired. Using only the $(0,0)$ and $(1,0)$  $(l,m)$ modes is the standard two-term approximation of the distribution function~\cite{hagelaar2005solving, COMSOL1998}. 

\subsubsection{Advection operator}
\label{subsubsec:advection_operator}
For the velocity space $z$-aligned electric field we have $\vect{E} = E \vect{\hat{e}_z} = E \of{\cos(\vtheta) \vect{\hat{e}_r} - \sin(\vtheta)\vect{\hat{e}_\theta}}$. Thus, in spherical coordinates, the advection term is given by  
\begin{equation}
    \vect{E} \cdot \nabla_{\vect{v}} f = E \of{\cos(\vtheta) \partial_v f - \sin(\vtheta) \frac{1}{v} \partial_{\vtheta}f }. \label{eq:adv_term}
\end{equation} Substituting \Cref{eq:f_expansion} in \Cref{eq:adv_term} and using associated Legendre polynomials recurrence relations, we obtain the strong from given by \Cref{eq:adv_term_a}. 
\begin{multline}
\vect{E} \cdot \nabla_{\vect{v}} f = E 
\sum_{klm} f_{klm} 
\Bigg(
\left( A_M\of{l,m} \frac{d}{d\vr}\phi_{k}\of{v} 
+ A_D\of{l,m} \frac{1}{\vr}\phi_{k}\of{v} \right)
Y_{(l-1)m}\of{\vtheta,\vphi} 
\\
+
\left( B_M\of{l,m} \frac{d}{d\vr}\phi_{k}\of{v} 
+ B_D\of{l,m} \frac{1}{\vr}\phi_{k}\of{v} \right)
Y_{(l+1)m}\of{\vtheta,\vphi} 
\Bigg). \label{eq:adv_term_a}
\end{multline}
The corresponding weak form for a given test function $\phi_{p}\of{v} Y_{qs}\of{\vtheta, \vphi}$ is specified in \Cref{eq:adv_ws} below.
\begin{multline}
    \myint_{\reals^+} \myint_{S^2} \Big(\vect{E}\cdot \nabla_{\vect{v}} f \Big) \phi_{p}\of{v} Y_{qs}\of{\theta,\phi} 
    \vr^2  \diff{\omega_v} \diff{\vr}\\
    = E 
    \myint_{\mathbb{R}^+} 
    \sum_{k}
    \Bigg(
    f_{k(q+1)s} 
    \left( A_M\of{q+1,s} \frac{d}{d\vr}\phi_{k}\of{v} 
    + A_D\of{q+1,s} \frac{1}{\vr}\phi_{k}\of{v} \right)
    \\
    + f_{k(q-1)s} \left( B_M\of{q-1,s} \frac{d}{d\vr}\phi_{k}\of{v} 
    + B_D\of{q-1,s} \frac{1}{\vr}\phi_{k}\of{v} \right)
    \Bigg)
    \phi_{p}\of{v}
    \vr^2 \diff{\vr}. \label{eq:adv_ws}
\end{multline}
In the above, $A_M, A_D, B_M$ and $B_D$ are coefficients indexed by $l,m$ (see \ref{subsec:advection}).

\subsubsection{Electron-heavy collisions}
\label{subsubsec:electron_heavy_collisions}
We consider binary reactions of the type $X + e \rightarrow \tilde{X} + e$, where $X$ denotes the pre-collision heavy particle species, and $\tilde{X}$ denotes the post-collision heavy particle species, i.e., excited, ionized, or unchanged (for elastic collisions). Let $\vect{v}_e$ and $\vect{v}_0$ be the electron and heavies velocities respectively. Let $f_{0}(\vect{v}_0)$ be the distribution function of the heavies and $f(\vect{v}_e)$ be the distribution function of the electrons.  Let us denote the maps from pre-collisional velocities $\vect{v}_e$, $\vect{v}_0$ to post-collisional velocities as $\vect{v}_e^\text{post} = \vect{v}_e^\text{post} \of{\vect{v}_e, \vect{v}_0, \vect{\omega}}$ and $\vect{v}_0^\text{post} = \vect{v}_0^\text{post} \of{\vect{v}_e, \vect{v}_0, \vect{\omega}}$ where $\vect{\omega} \in S^2$ is the solid scattering angle. The electron-heavy collision operator is derived by considering the number of electrons lost and gained for a specified electron velocity $\vect{v}_e$. Let us first define the \emph{collision probability kernel} $B(\vect{v}_e,\vect{v}_0,\vect{\omega})$ which gives a probability measure for the occurrence of electron-heavy collision with $(\vect{v}_e,\vect{v}_0, \vect{\omega})$. \Cref{eq:c_en_loss}~describes loss of electrons at velocity $\vect v_e$, while \Cref{eq:c_en_gain}~describes the creation of electrons at velocity $\vect v_{e}$ due to underlying collisional process.  The strong form of the electron-heavy binary collision operator is given in~\Cref{eq:c_en}.

%
\begin{align}
C^{-}_{en}(f,f_0) &= \myint_{\reals^3} \myint_{\reals^3} \myint_{S^2} 
B\of{\vect{v}_e^\prime, \vect{v}_0^\prime, \vect{\omega}} 
f\of{\vect{v}_e^\prime} f_0\of{\vect{v}_0^\prime} 
\delta\of{\vect{v}_e^\prime - \vect{v}_e} 
\diff{\vect{v}_0^\prime} \diff{\vect{v}_e^\prime} \diff{\vect{\omega}}, \label{eq:c_en_loss}
\\
C^{+}_{en}\of{f,f_0} &= \myint_{\reals^3} \myint_{\reals^3} \myint_{S^2} 
B\of{\vect{v}_e^\prime, \vect{v}_0^\prime, \vect{\omega}} 
f\of{\vect{v}_e^\prime} f_0\of{\vect{v}_0^\prime} 
\delta\of{\vect{v}_e^\text{post}\of{\vect{v}_e^\prime, \vect{v}_0^\prime, \vect{\omega}} - \vect{v}_e} 
\diff{\vect{v}_0^\prime} \diff{\vect{v}_e^\prime} \diff{\vect{\omega}} \label{eq:c_en_gain},
\\
C_{en}\of{f,f_0} &= C^{+}_{en}\of{f,f_0} - C^{-}_{en}\of{f,f_0}. \label{eq:c_en}
\end{align}
Note that the electron-heavy collision operator is a function of $\vect{v}_e$. For a given test function $\Phi(\vect{v}_e)$, the weak form of the electron-heavy collision operator is given in~\Cref{eq:wf_c_en}. Note that \Cref{eq:c_en} and the corresponding weak form given in~\Cref{eq:wf_c_en} describe the most general form of the electron-heavy binary collision operator.
\begin{align}
\myint_{\reals^3} C^{-}_{en} (f,f_0) \Phi\of{\vect{v}_e} \diff{\vect{v}_e} 
&=
\myint_{\reals^3} \myint_{\reals^3} \myint_{S^2} 
B\of{\vect{v}_e, \vect{v}_0, \vect{\omega}} 
f\of{\vect{v}_e} f_0\of{\vect{v}_0} 
\Phi\of{\vect{v}_e} 
\diff{\vect{v}_e} \diff{\vect{v}_0} \diff{\vect{\omega}}, \label{eq:wf_c_en_loss}
\\
\myint_{\reals^3} C^+ (f,f_0)\Phi\of{\vect{v}_e} \diff{\vect{v}_e} 
&= 
\myint_{\reals^3} \myint_{\reals^3} \myint_{S^2} 
B\of{\vect{v}_e, \vect{v}_0, \vect{\omega}} 
f\of{\vect{v}_e} f_0\of{\vect{v}_0} 
\Phi\of{\vect{v}_e^\text{post}\of{\vect{v}_e, \vect{v}_0, \vect{\omega}}} 
\diff{\vect{v}_0} \diff{\vect{v}_e} \diff{\vect{\omega}}, \label{eq:wf_c_en_gain}
\\
\myint_{\reals^3} C_{en}(f,f_0) \Phi\of{\vect{v}_e} \diff{\vect{v}_e} 
&=
\myint_{\reals^3} \myint_{\reals^3} \myint_{S^2} 
B\of{\vect{v}_e, \vect{v}_0, \vect{\omega}} 
f\of{\vect{v}_e} f_0\of{\vect{v}_0} 
\left(
\Phi\of{\vect{v}_e^\text{post}\of{\vect{v}_e, \vect{v}_0, \vect{\omega}}} 
- \Phi\of{\vect{v}_e} 
\right)
\diff{\vect{v}_0} \diff{\vect{v}_e} \diff{\vect{\omega}}. \label{eq:wf_c_en}
\end{align}

In low gas temperatures, e.g., glow discharge devices, the velocities of the heavies are negligible compared to the electron velocities. In that case \Cref{eq:wf_c_en} can be further simplified by setting $f_0\of{\vect{v_0}} = n_0 \delta\of{\vect{v_0}}$ where $\delta$ denotes the Dirac delta. The collision kernel $B$ is modeled using experimental collisional cross-section data~\cite{pitchford2017lxcat}. With the above, we write, $B\of{\vect{v}_e, \vect{0}, \vect{\omega}} = \norm{\vect{v}_e} \sigma\of{\norm{\vect{v}_e}, \vect{\omega}}$, where $\sigma$ denotes the differential cross-section for the underlying electron-heavy collisional process. Under the above conditions, we can simplify \Cref{eq:wf_c_en} as in \Cref{eq:wf_c_en_n0}. We remark that, even in LTPs, $f_0\of{\vect{v_0}} = n_0 \delta\of{\vect{v_0}}$ is not always the case. If the background heavy temperature is higher, as for example in torch devices, $f_0\of{\vect{v_0}}$ is assumed to be a Maxwellian distribution and is integrated out from \Cref{eq:wf_c_en}.
\begin{equation}
    \myint_{\reals^3} C_{en}(f) \Phi\of{\vect{v}_e} \diff{\vect{v}_e} 
    =
    n_0\myint_{\reals^3} \myint_{S^2} 
    \norm{\vect{v}_e} \sigma \of{\norm{\vect{v}_e},\omega} 
    f\of{\vect{v}_e} 
    \left(
    \Phi\of{\vect{v}_e^\text{post}\of{\vect{v}_e, \vect{0}, \vect{\omega}}} 
    - \Phi\of{\vect{v}_e} 
    \right)
    \diff{\vect{v}_e} \diff{\vect{\omega}}. \label{eq:wf_c_en_n0}
\end{equation}

The collision operator in \Cref{eq:wf_c_en_n0} is a 5d integral that captures electron-heavy collisions. With the isotropic scattering assumption, we can write $\sigma(\norm{\vect{v}_e}, \vect{\omega}) = \frac{\sigma_{T}\of{\norm{\vect{v}_e}}}{4\pi}$ where $\sigma_T$ denotes the total cross-section. Total cross sections can be found in the LXCAT database~\cite{pitchford2017lxcat}. Using total cross-sections and the z-axis aligned with the $\vect{E}$ field,  the spatially homogeneous BTE preserves azimuthal symmetry. This symmetry can be used to further simplify the collision operator given in~\Cref{eq:wf_c_en_n0} by analytically integrating out the angular integrals.

\textbf{Simplified collision operator:} Let the scattering solid angle be  $\vect{\omega} = (\chi, \phi)$ respectively. Let $\Phi_{klm}, \Phi_{pqs}$ denote the trial and test basis functions. For an incoming electron velocity $\vect{v}=(v,\vtheta, \vphi)$, let $\vect{u}=(u, \utheta, \uphi)$ be the post-collision velocity (i.e., $\vect{u} = \vect{v}^{\text{post}} \of{\vect v, \vect{0}, \vect{\omega}}$). The discretized $C_{en}$ operator in the matrix form can be written as in \Cref{eq:c_en_mat}. 
\begin{equation}
    n_0{[\vect C_{en}]}^{pqs}_{klm} = n_0 \myint_{\reals^+} \myint_{S^2} \myint_{S^2} 
    v^2 \sigma\of{\vect{v},\vect{\omega}} \phi_{k}\of{v} Y_{lm}\of{\vtheta,\vphi}
    \of{\phi_{p}\of{u} Y_{qs}\of{\utheta,\uphi} -\phi_{p}\of{v} Y_{qs}\of{\vtheta,\vphi}} \diff{\omega_v} \diff{\omega}\diff{v}. \label{eq:c_en_mat}
\end{equation} For the post-collision velocity $\vect{u}$, we can show that $\cos(\utheta) = \cos(\vtheta) \cos(\chi) + sin(\vtheta) \sin(\chi) \cos(\vphi-\phi)$~\cite{allis1956motions}. Further, we can integrate the angular integrals analytically using the spherical harmonics addition theorem (see \ref{subsec:collop_simplifications}). Under azimuthal symmetry with isotropic scattering the collision operator becomes
\begin{equation}
    n_0 {[\vect C_{en}]}^{pq}_{kl} = n_0 \myint_{\reals^+} v^2 \sigma_T\of{v} \phi_{k}\of{v} \delta_{ql} \of{\phi_{p}\of{u} \delta_{q0}  -\phi_{p}\of{v}} \diff{v}. \label{eq:c_en_mat_1d}
\end{equation}
A similar expression can be derived for non-zero heavy temperature, where we assume heavy species distribution function is Maxwellian $f_{0}\of{\vect{v}} = \frac{n_0}{\of{\sqrt \pi v_{th}}^3} \exp\of{\of{-v/v_{th}}^2}$ where $v_{th}=\sqrt{2k_B T_0/m_0}$ is the thermal velocity, $n_0$ is the number density, and $m_0$ is the particle mass of the heavy species. For non-zero heavy temperature \Cref{eq:c_en_mat_1d} gets an additional correction term, given by \Cref{eq:c_T0_mat_1d}.
\begin{equation}
    n_0 T_0 {[\vect C_{T}]}^{pq}_{kl} = \frac{n_0 T_0 k_B}{m_0} \delta_{q0} \myint_{\reals^+} v^3 \sigma_T\of{v} \partial_v \phi_p\of{v} \partial_v \phi_k\of{v} \diff{v}. \label{eq:c_T0_mat_1d}
\end{equation} 
For a plasma with multiple collisions (i.e., elastic, excitation, ionization, step-excitation), we can define an effective collision operator as $C_{en}^{\text{effec.}} = \sum_{i\in {\text{collisions}}}$ $n_i C_{en} \of{\sigma_i} + n_0 T_0 C_{T}\of{\sigma_0}$ where $0$ index denotes the ground state heavy species and $\sigma_0$ denotes the total cross-section for elastic collisions.


\subsubsection{Electron-electron Coulomb interactions}
\label{subsubsec:coulomb_collisions}
Coulomb interactions between charged particles are modeled using the inverse square force formulation proposed by~\cite{rosenbluth1957fokker}. We focus on electron-electron Coulomb interactions modeled using a Fokker-Plank equation specified by \Cref{eq:FP_sf}, where $\ln\Lambda$ is the Coulomb logarithm, $h\of{\vect{v}}, g\of{\vect{v}}$ denote the Rosenbluth potentials, $H$ denotes the Hessian of $f$, $T_e, \epsilon_0$, $m_e$ denote electron temperature, vacuum permittivity, and electron mass in SI units (i.e., K, F/m, Kg) respectively.


\begin{align}
    \Gamma_a ^{-1} C_{ee}(f,f) = -\nabla \cdot (f \nabla h ) + \frac{1}{2} H : \of{f H g}, \text{ where } \Gamma_a = \frac{e^4 \ln \Lambda}{4\pi \epsilon_0^2 m_e^2},\ \Lambda = \frac{12 \pi (\epsilon_0 k_B T_e)^{3/2}}{e^3 n_e^{1/2}}, \label{eq:FP_sf} \\
     \Delta h\of{\vect{v}} = -8\pi f\of{\vect{v}}, \text{ and } \Delta^2 g\of{\vect{v}} = -8\pi f\of{\vect{v}} \label{eq:RP_ee}.  
\end{align} The weak form of the electron-electron Coulomb collision operator can be written as \Cref{eq:FP_ws}.
\begin{align}
    \myint_{\reals^3} C_{ee}(f,f) \Phi\of{\vect{v}} \diff{\vect{v}} &= \Gamma_a \myint_{\reals^3} \Phi\of{\vect{v}} \of{-\nabla \cdot (f \nabla h ) + \frac{1}{2} H : \of{f H g}} \diff{\vect{v}} \text{}\nonumber \\
    & = \Gamma_a \myint_{\reals^3} \of{\nabla \psi\of{\vect{v}} \cdot (f \nabla h )  + \frac{1}{2} \int_{\vect{v}} H\psi\of{\vect{v}} : \of{f H g} } \diff{\vect{v}}. \label{eq:FP_ws}
\end{align}
Using the free space harmonic and bi-harmonic Green's functions, $h$ and $g$ can be explicitly written as moments of $f$. Accounting for azimuthal symmetry, we can expand the potentials $h\of{\vect{v}}$ and $g\of{\vect{v}}$ in spherical harmonics as follows:
\begin{equation}
    h\of{\vect{v}} = \sum_{s} h_{s}\of{v} Y_{s0}\of{\vtheta, \vphi} \text{ and } g\of{\vect{v}} = \sum_{s} g_{s} \of{v} Y_{s0}\of{\vtheta, \vphi}. \label{eq:potential_expansion}
\end{equation}
The coefficients for the expansion (i.e., functions of radial coordinate $v$) can be computed using moments of the distribution function $f$ as shown in \Cref{eq:hs_v} and \Cref{eq:gs_v}. In the above $f_s$ denotes the corresponding polar mode in the expansion $f(\vect{v}) = \sum_{s} f_s\of{v} Y_{s0}\of{\vtheta, \vphi}$.
%
\begin{align}
    h_s\of{v} &= \frac{8\pi}{2s + 1} \of{\int_{0}^{v} dv^\prime \frac{{v^\prime}^{s+2}}{v^{s+1}} f_s\of{v^\prime} +  \int_{v}^{\infty} dv^\prime \frac{{v}^{s}}{{v^\prime}^{s-1}} f_s\of{v^\prime}} \label{eq:hs_v} ,\\
    g_s\of{v} &= -\frac{4\pi}{(4s^2-1)} \of
	{
		\int_{0}^{v} \diff{v^\prime} f_s\of{v^\prime} \frac{{v^\prime}^{s+2}}{v^{s-1}} \of{1 - \of{\frac{s-1/2}{s+3/2}}\of{\frac{{v^\prime}^2 }{v^2}}} } + \nonumber\\
	&-\frac{4\pi}{(4s^2-1)} \of{
		\int_{v}^{\infty} \diff{v^\prime} f_s\of{v^\prime} \frac{{v}^{s}}{{v^\prime}^{s-3}} \of{1 - \of{\frac{s-1/2}{s+3/2}}\of{\frac{{v}^2 }{{v^\prime}^2}}}
    }.\label{eq:gs_v}
\end{align}
Further, we expand  $f_s\of{v}, h_s\of{v}$ and $g_s\of{v}$, in the radial direction the B-spline basis, i.e., $f_s\of{v} = \sum_{r} f_{rs} \phi_r\of{v}$, $h_s\of{v} = \sum_{r} h_{rs} \phi_r\of{v}$ and $g_s\of{v} = \sum_{r} g_{rs} \phi_r\of{v}$. The above expansion allows us to precompute operators $H^{rs}_{kl}$ and $G^{rs}_{kl}$ that can be used to compute the Rosenbluth potentials in B-spline and spherical basis expansion, i.e.,  $h_{rs} = \sum_{kl} H^{rs}_{kl} f_{kl}$ and $g_{rs} = \sum_{kl} G^{rs}_{kl} f_{kl}$. Therefore, we can write the discretized Coulomb operator as a tensor given by \Cref{eq:cc_ee_tensor}, where $\Phi_{kl}\of{\vect{v}}=\phi_{k} Y_{l0}\of{\vtheta,\vphi}$ and $\Phi_{rs}\of{\vect{v}}=\phi_{r} Y_{s0}\of{\vtheta,\vphi}$ denote the trial and test basis functions. 
\begin{equation}
    [\vect C_{ee}]^{pq}_{kl, rs} =  \myint_{\reals^3} \of{\nabla \Phi_{pq} \cdot (\Phi_{kl} \nabla \Phi_{rs} )  + \frac{1}{2} H\Phi_{pq} : \of{\Phi_{kl} H \Phi_{rs}} } \diff{\vect{v}}. \label{eq:cc_ee_tensor}
\end{equation}
Then the weak form of the Coulomb operator is given by \Cref{eq:cc_ee_wf_tensor}.
\begin{equation}
\myint_{\reals^3} C_{ee}(f,f) \Phi_{pq}\of{\vect{v}} \diff{\vect{v}} = \Gamma\of{f_{kl}} [C_{ee}]^{pq}_{kl,rs}(f_{kl}, f_{rs}) \label{eq:cc_ee_wf_tensor}.
\end{equation}

In spherical coordinates, the divergence and Hessian operators have additional correction terms. To simplify their implementation, we have developed SymPy~\cite{meurer2017sympy} based Python code generation module to assemble Coulomb collision tensor for a specified number of l-modes. The computation of the $\nabla$ and the Hessian operators in spherical coordinates are given in \ref{subsec:ee_collisions}. Once the symbolic expressions are assembled, angular integrals are performed analytically using symbolic integration. For certain $(q,s,l)$ combinations, angular integrals are zero, thus reducing the overall computation and storage cost of Coulomb tensor assembly.

\subsubsection{Boundary conditions}
The discretized operators above involve integrals on unbounded velocity space. In spherical coordinates this reduces to invoking boundary conditions in the radial direction. We assume that for all time $t$, $\lim_{\norm{v}\rightarrow \infty} f(t,\vect{v}) \rightarrow 0$. Numerical evaluation of the velocity space integrals is handled with domain truncation with specified maximum energy $\varepsilon_{max}$. Currently, $\varepsilon_{max}$ has to be specified as a priory based on the maximum energy allowed on the grid for given input conditions (i.e., electric field, ionization degree, and heavy temperature). There are several heuristics to determine the $\varepsilon_{max}$. We use Maxwellian EDF at approximate electron temperature to determine the $\varepsilon_{max}$.



%

\subsection{Steady-state and transient solutions}
\label{subsec:ss_and_transient_sol}
In this section we discuss the solvers for \cref{eq:bte}. These are based on its discretized form.  Let $N_r$ and $N_l$ denote the number of radial B-splines and spherical modes used in  the discretization of \Cref{eq:bte}. The discretized spatially homogeneous BTE is given by \Cref{eq:discretized_bte}, where $N_v=N_r N_l$, $\vect{f}\in \reals^{N_v}$, $\vect{C}_{en}\in \reals ^{N_v \times N_v}$, $\vect{A}_v \in \reals ^{N_v \times N_v}$ and $C_{ee} \in \reals^{N_v \times N_v \times N_v}$.
\begin{equation}
    \partial_t \vect{f} = (\vect{C}_{en} + E \vect{A}_v) \vect{f} + \vect{C}_{ee}\of{\vect{f}, \vect{f}} \label{eq:discretized_bte}.
\end{equation}
$\vect{E}$ causes electron acceleration while electrons undergo energy loss due to collisions. We limit our attention to electron-heavy binary collisions, ignoring three-body recombination collisions. With the above, we only have an electron production mechanism without any electron loss. Therefore, we need to consider the normalized EDF when considering steady-state solutions. We can write the evolution equation for the normalized EDF as in \Cref{eq:normalize_evol}. By substituting \Cref{eq:discretized_bte} in \Cref{eq:normalize_evol}, we can write the discretized dynamical system for the normalized EDF as in \Cref{eq:normalized_evol_1}.
\begin{equation}
    \partial_t {\hat{f}} = \frac{1}{n_e} \partial_t f - \hat{f} \of{\frac{\partial_t n_e }{n_e}} \text{ where } n_e(t) = \int_{\reals^3} f(t,\vect{v}) \diff{\vect{v}} \text{ and } \hat{f}\of{t,\vect{v}} = \frac{1}{n_e(t)} f\of{t,\vect{v}}. \label{eq:normalize_evol}
\end{equation}
\begin{equation}
    \partial_t \vect{\hat{f}} = (\vect{C}_{en} + E \vect{A}_{v}) \vect{\hat{f}} + n_e \vect{C}_{ee}\of{\vect{\hat f}, \vect{\hat f}} - {\vect{\hat f}} \of{\frac{\partial_t n_e }{n_e}}. 
    \label{eq:normalized_evol_1}
\end{equation} 
For the term $\frac{\partial_t n_e}{n_e}$ by integrating both sides of \Cref{eq:bte}, we can derive \Cref{eq:mass_growth_continuous}. In the above, using integration by parts, we can show $ \int_{\reals^3} \vect{E}\cdot \nabla_{\vect{v}} f \diff{\vect{v}} =0 \text { and } \int_{\reals^3} C_{ee}\of{f,f} \diff{\vect{v}} = 0$. 
\begin{align}
    \frac{\partial_t n_e}{n_e} &= \frac{1}{n_e}\int_{\reals^3} \of{C_{en}\of{f} + C_{ee}\of{f,f} + \frac{q_e\vect{E}}{m_e}\cdot\nabla_{\vect{v}} f} \diff{\vect{v}} = \frac{1}{n_e}\int_{\reals^3} C_{en}\of{f} \diff{\vect{v}} =  \int_{\reals^3} C_{en}({\hat{f}}) \diff{\vect{v}} 
    \label{eq:mass_growth_continuous}
\end{align} 
For the discretized system we can write the mass growth term $\frac{\partial_t n_e}{n_e} = \vect{u}^T \vect{C}_{en} \vect{\hat{f}}$ where $[\vect{u}]_{kl} = \int_{\reals^3} \phi_k\of{v} Y_{l0}\of{\vtheta, \vphi} \diff{\vect{v}}$. With that, we can write the final discretized equation for the normalized distribution function as \Cref{eq:normalized_evol_2} with the constraint $\vect{u}^T \vect{\hat{f}} = 1$.
\begin{subequations}
    \begin{align}
        &\partial_t {\vect{\hat{f}}} = (\vect{C}_{en} + E \vect{A}_v) \vect{\hat{f}} + n_e \vect{C}_{ee}(\vect{\hat{f}}, \vect{\hat{f}}) - \mu(\vect{\hat{f}}) \vect{\hat{f}} \text{ where } \mu(\vect{\hat{f}}) = \vect{u}^T \vect{C}_{en} \vect{\hat{f}},\label{eq:normalized_evol_2} \\
        &\vect{u}^T \vect{C}_{en}\vect{\hat{f}}= 1. \label{eq:normalized_constraint}
    \end{align}\label{eq:normalized_bte}
\end{subequations}

In our transient solver we need to evolve \Cref{eq:normalized_evol_2} with the constraint $\vect{u}^T \vect{\hat{f}} =1$. We do so using a projection scheme to numerically enforce the constraint above. Let $\vect{p}=\frac{\vect u}{\norm{\vect{u}}_2}$. Let the QR factorization $(\vect{I}-\vect{p} \vect{p}^T)=\vect{Q}_{N_v\times (N_v-1)} \vect{R}_{(N_v-1)\times N_v}$. Then we can decompose $\vect{\hat{f}} = \vect{f}_1 + \vect{Q} \vect{\bar{f}}$ such that $\vect{u}^T \vect{f}_1 = 1$ and $\vect{u}^T \vect{Q}\vect{\bar{f}}=0$ where $\vect{\bar{f}}\in \reals^{N_v-1}$. Let $\vect{f}_1= \frac{\vect{u}}{\norm{\vect{u}}_2^2}$. Note that $\partial_t \vect{\hat{f}} = \partial_t (\vect{f}_1 + \vect{Q}\vect{\bar{f}}) = \vect{Q}\partial_t \vect{\bar{f}}$. Therefore, we can write the evolution of $\vect{\bar{f}}$ as \Cref{eq:f_bar_evol} with an initial condition given by
$\vect{\bar{f}}_{t=0} = \vect{R} \vect{f}_{t=0}$.
\begin{equation}
    \partial_t \vect{\bar{f}} = \vect{Q}^T (\vect{C}_{en} + E \vect{A_v}) \vect{\hat{f}} + n_e \vect{Q}^T \vect{C}_{ee}(\vect{\hat{f}}, \vect{\hat{f}}) -\mu(\vect{\hat{f}}) \vect{Q}^T \vect{\hat{f}} \label{eq:f_bar_evol}
\end{equation} 

\textbf{Steady state solutions}: For a static $\vect{E}>0$ field, the dynamical system given by~\Cref{eq:normalized_bte} should reach a steady-state solution. This solution can be  computed using  Newton's method to solve \Cref{eq:rhs_bte_ss}. The Jacobian of \Cref{eq:rhs_bte_ss} is given by \Cref{eq:rf_jacobian}.
\begin{align}
    R(\vect{\bar{f}})=\vect{Q}^T (\vect{C}_{en} + E \vect{A}_v) \vect{\hat{f}} + n_e \vect{Q}^T \vect{C}_{ee}(\vect{\hat{f}}, \vect{\hat{f}}) -\mu(\vect{\hat{f}}) \vect{Q}^T \vect{\hat{f}} = \vect{0}. 
    \label{eq:rhs_bte_ss} \\
\frac{d R}{d \vect{\bar{f}}}\of{\vect{\hat{f}}} = \vect{Q}^T \of{\vect{C}_{en} + E \vect{A}_v + n_e \vect{C}_{ee}(\cdot, \vect{\hat{f}}) + n_e \vect{C}_{ee}(\vect{\hat{f}}, \cdot) - \vect{u}^T \mu(\vect{\hat{f}}) \vect{Q}^T} \vect{Q}.  \label{eq:rf_jacobian}
\end{align}

    

\textbf{Transient solutions}:
For time-harmonic $\vect{E}\of{t}$, the dynamical system given in~\Cref{eq:normalized_bte} would result in time-periodic EDF solutions at the steady-state. The above can be computed through time integration of~\Cref{eq:normalized_bte}. We use an implicit backward Euler scheme to time march. 

\textbf{Complexity analysis}:
We summarize the complexity analysis for the developed Eulerian solver. Let $N_r$, $N_l$ be the number of basis functions used in the radial and angular directions. For a given $N_r$, $N_l$, and a specified energy range $\varepsilon_{max}$, we can pre-compute all the discretized operators needed for the solve. The above only depends on the grid parameters $N_r$, $N_l$, and $\varepsilon_{max}$ and can be used for multiple solves, given that grid parameters remain unchanged. Assuming the $\mathcal{O}(1)$ evaluation of the basis functions, we can write the computational complexity of the electron-heavy collision operator given in \Cref{eq:c_en_mat_1d} as $\mathcal{O}(N_r^2)$. Due to compact support of B-splines, each radial polynomial $p$ interacts with splines ranges in $[\max(0,p-(d+2)), \min(N_r,p + (d+2))]$ where $d$ denotes the order of the B-spline basis. With the above, we can write the computational complexity of the discretized Coulomb collision operator given in \Cref{eq:cc_ee_tensor} as $\mathcal{O}(N_r N_l^3)$. Similarly, we can write the computational complexity for $\vect{v}$-space advection operator given in \Cref{eq:adv_ws} and the mass matrix as $\mathcal{O}(N_r)$. Both steady-state and transient solvers require inner liner solve for the Newton iteration. For the above, we use LU factoring-based direct solve. \Cref{tab:complexity} summarizes the key operations and their complexities for the Eulerian solver. The electron-electron Coulomb interactions and the Jacobian factorization dominate the overall cost of the Newton iteration. 

\begin{table}[!tbhp]
	\centering
	\begin{tabular}{|c|c|c|}
		\hline 
		Operation / operator(s) & Dimension & Action \\
		\hline
		$\vect{C}_{ee}$ & $N_v \times N_v \times N_v$ & $\mathcal{O}(2 N_v^3)$ \\
		$\vect{C}_{en}$, $\vect{A}_v$ & $N_v \times N_v$ & $\mathcal{O}(N_v^2)$\\
		Jacobian (assembly) & $N_v \times N_v$ & $\mathcal{O}(N_v^3)$ \\
		Jacobian (LU solve) & $N_v \times N_v$ & $\mathcal{O}(\frac{2}{3} N_v^3)$\\
		\hline
	\end{tabular}
\caption{Computational complexity of key operations for the Eulerian BTE solver. \label{tab:complexity}}
\end{table}

\textbf{Implementation}: The developed solver is implemented in Python~\cite{python3} programming language. Once all the operators are assembled, the assembled operators' action (i.e., matrix or tensor contractions) is implemented using standard linear algebra libraries. We rely on  \texttt{NumPy}~\cite{numpy} and \texttt{CuPy}~\cite{cupy_learningsys2017} libraries for implementation of the Eulerian solver in CPU and GPU architectures. \texttt{NumPy} library routines provide an interface to BLAS~\cite{BLAS} linear algebra routines for CPUs. \texttt{CuPY} provides an interface for GPU accelerated linear algebra routines through \texttt{cuBLAS}~\cite{clblas} and \texttt{rocBLAS}~\cite{rocBLAS} libraries. 



	\subsection{Direct simulation Monte-Carlo solver}\label{sec:dsmc}
For the direct simulation Monte-Carlo (DSMC) approach, we use  Lagrangian particle tracking electron kinetics. In particular, we track $N(t)$ particles, where each particle represents  $\frac{n_e}{N}$ real electrons. At each timestep, particles undergo an advection step and a collision step. Advection in velocity-space is done using a forward Euler time stepping scheme. Our electron collision method is based on~\cite{vahedi1995monte} for electron-heavy collisions and \cite{wang2008particle} for the Coulombic interactions. These collision steps are detailed below.

\subsubsection{Electron-heavy collisions}
For the modeling of electron-heavy collisions, unlike~\cite{vahedi1995monte}, we do not use the null collision method. Our numerical experiments found it to introduce errors without providing notable performance speedup. At each timestep, the collision step is performed as described in \cref{alg:collision_dsmc}.
\begin{algorithm}[!tbhp]
    \caption{Overview: DSMC collision step}\label{alg:collision_dsmc}
    \begin{algorithmic}
    \Require Gas density $n_0$, particle simulators $S_t$ (i.e., at time $t$), and timestep size $\Delta t$,
    \Ensure $S_{t + \Delta t}$
    \For{$s_i$ in $S_t$ }
        \State $\varepsilon_i \leftarrow \frac{1}{2} m_e v_i^2$ \Comment{compute particle energy}
        \State $P_i \leftarrow 1-\exp(-\Delta t \norm{v_i}_2 n_0 \sum_{j}\sigma_j(v_i))$ \Comment{total collision probability, where $j$ denotes the collision type}
        \State $R_1 \leftarrow \text{Random}(0,1)$ \Comment{randomization of collision process}
        \If {$R_1 \leq P_i$}
            \State Select collision type $j$ with probability $\frac{\sigma_j}{\sum_j \sigma_j}$
            \State $\cos{(\chi)} \leftarrow \text{Random}(-1,1)$, $\phi \leftarrow \text{Random}(0,2\pi)$ \Comment{randomized scattering angles}
            \State $v_{s_i}^{t+\Delta t} \leftarrow C_j(v_{s_i}^{\Delta t},\chi,\phi)$ \Comment{Execute collision type $j$} 
        \EndIf
    \EndFor
    \State \Return 
\end{algorithmic}
\end{algorithm}
We now discuss the details of the collision process. We detail the case for elastic collisions; other collision types have nearly identical collision scattering angles with different energy thresholds. First, we define the post-collision velocity unit vector
\begin{equation} \label{dsmc:colldirn}
    \mathbf{v_{\mathrm{scat}}} =\mathbf{v_{\mathrm{inc}}} \cos{\chi} + \mathbf{v_{\mathrm{inc}}} \times \mathbf{i} \frac{\sin{\chi}\sin{\phi}}{\sin{\theta}} + \mathbf{v_{\mathrm{inc}}} \times (\mathbf{i} \times \mathbf{v_{\mathrm{inc}}}) \frac{\sin{\chi}\cos{\phi}}{\sin{\theta}},
\end{equation} 
 where $\chi$ is the scattering polar angle; $\cos{\chi}$ is drawn uniformly from $\big[-1,1\big]$. $\phi$ is the azimuthal angle and is drawn uniformly from $\big[0,2\pi\big]$. $\theta$ is defined as $\cos{\theta} = \mathbf{v_{\mathrm{inc}}} \cdot \mathbf{i}$ where $\mathbf{v_{\mathrm{inc}}}$ denotes the unit vector along the collision incident (i.e., the relative velocity between electron and heavy particles). 
We note that many codes consider anisotropic scattering angles $\chi$ that depend on the incoming velocity vector. 
To maintain consistency between this DSMC code and the presented Eulerian approach, we use isotropic scattering assumption and choose $\chi$ as described in \cref{alg:collision_dsmc}.

Now that we have defined the post-collision direction of a particle, we must determine the post-collision speed. The post-collision speed of an electron depends on the underlying electron-heavy collisional process. For example, elastic collisions have relative energy loss term $\Delta \epsilon = \frac{2 m_e}{m_0}$ where $m_0$ represents the heavy species particle mass. In contrast, for ionization collisions, we have an energy threshold of $\varepsilon_{ion}>0$, and assuming equal energy sharing, the post-collision energy of each particle $\epsilon^\prime$ is given by $\varepsilon^\prime = \frac{\varepsilon - \varepsilon_{\mathrm{ion}}}{2}$.


\subsubsection{Coulomb collisions}
We follow ~\cite{wang2008particle} for the electron-electron Coulomb interactions. At each timestep, all $N$ particles are randomly grouped into $\frac{N}{2}$ pairs, and a collision is performed between these two particles. For each pair of particles $(i,j)$, we first calculate the relative velocity vector $u = v_i - v_j$. We draw $\phi$ randomly such that $\phi \in [0,2\pi]$ and $\delta$ from a Gaussian distribution with zero mean and variance as given by
\begin{equation}
    \langle \delta^2 \rangle = \frac{q_e^4 n_e \Delta t \log\Lambda}{2 \pi \epsilon_0^2 m_e^2 u^3}, \label{eq:dsmc_cc_varience}
\end{equation}
where $q_e$ is the unit charge; $n_e$ is the electron density; $\Delta t$ is the timestep; $\log\Lambda$ is the Coulomb logarithm; $\epsilon_0$ is the vacuum permittivity.
%
We define $\sin{\theta} = \frac{2\delta}{1+\delta^2}$ and $\cos{\theta} = 1- \frac{2\delta^2}{1+\delta^2}$. With the above we can write the velocity component updates for electron-electron Coulomb interactions by
\begin{equation}
    \begin{split}
        \Delta u_x = \frac{u_x u_z \sin(\theta)\cos(\phi)}{u_p} -\frac{u_y \big|u\big| \sin(\theta) \sin(\phi)}{u_p} - u_x(1-\cos(\theta)), \\
        \Delta u_y = \frac{u_y u_z \sin(\theta)\cos(\phi)}{u_p} +\frac{u_x \big|u\big| \sin(\theta) \sin(\phi)}{u_p}- u_y(1-\cos(\theta)), \ 
        \Delta u_z = -u_p \sin(\theta) \cos(\phi) - u_z(1-\cos(\theta)).
    \end{split} \label{eq:dsmc_vel_update}
\end{equation} Writing $\vect{\Delta u} = \big[\Delta u_x, \Delta u_y, \Delta u_z\big]$, the post collisional velocities $\vect v_i'$, $\vect v_j'$ are given by $\vect v_i' = \vect v_i + \frac{\vect{\Delta u}}{2}$ and $\vect v_j' = \vect v_j + \frac{\vect{\Delta u}}{2}$ respectively. 

\subsection{Implementation}

The DSMC code we use here is a 0-D adaptation of a 1-D particle-in-cell electron Boltzmann solver that is fully detailed in \cite{almgrenbell}. The code is built in Python~\cite{python3}, primarily relying on \texttt{CuPy}~\cite{cupy_learningsys2017} data structures. This allows us to take advantage of GPU accelerated functions such as \texttt{CuPy}'s random number generation. Most of the time stepping computation for particle evolution, including all collisions types and advection, is done using custom GPU kernels implemented in \texttt{Numba}~\cite{numba} and \texttt{PyKokkos}~\cite{pykokkos}.


\section{Results} \label{sec:results}
A detailed summary of the conducted experiments is presented in \Cref{tab:summary_experiments}. The presented results are organized as follows. Convergence results for the Eulerian and DSMC solvers are presented in \Cref{subsec:conv_pde} and \Cref{subsec:conv_dsmc} respectively. The verification with existing state-of-the-art \bolsig~code~\cite{hagelaar2005solving,hagelaar2015coulomb} is presented in \Cref{subsec:verification}. A detailed perturbation study on the high-order $l$-modes and corresponding relaxation time scales are presented in \Cref{subsec:perturbations}.

\begin{table}[!tbhp]
	\centering
	\begin{tabular}{|p{2cm}|p{14cm}|}
		\hline
		Experiment & Description \\
		\hline
		Table \ref{tab:self_convergence} & Self-convergence results for the electron distribution function and subsequent QoI computations performed by the developed Eulerian Boltzmann solver. \\
		\hline
		Table \ref{tab:ss_perfornance} & Summary of the overall solve cost for the steady-state Eulerian BTE solver. \\
		\hline
		Table \ref{tab:dsmc_performance} & Summary of the overall time to solution for the DSMC BTE solver for different electric field values. \\
		\hline
		Table \ref{tab:dsmc_self_conv2} & Self-convergence results for the DSMC solver for different Boltzmann simulation parameters \\
		\hline
		Table \ref{tab:errors_with_bolsig} & QoI relative errors for the Eulerian and DSMC Boltzmann solvers compared to  \bolsig\\
		\hline
		Figure \ref{fig:convergence_rate} & Rate of convergence for several selected runs from \Cref{tab:self_convergence}. The above shows the rate of convergence of the relative $L_2$ error (i.e., taking $N_r$=256 as the exact solution) of the electron energy density function (i.e., $f_0$) computed with fixed two-term approximation with increasing resolution in the radial direction with cubic B-splines. As expected, the $L_2$ norm shows fourth order convergence rate for the cubic B-splines.\\
		\hline
		Figure \ref{fig:ho_modes_convergence_wo_cc_terms} & Convergence of steady-state EDF radial components $f_l\of{\vr}$ with respect to the number of angular terms used in the approximation. The input parameters for the above run is given by $E/N$=100Td and $n_e/n_0=0$. \\
		\hline 
		Figure \ref{fig:hl_convergence_4term} & Computed higher order $l$-modes (i.e., $f_2$ and $f_3$ modes) using four-term approximation with increasing resolution in the radial coordinate for input parameters $E/n_0$=100Td, and $n_e/n_0=0$. Increased radial resolution helps to capture the high energy tails of the computed $l$-modes. \\
		\hline 
		Figure \ref{fig:bolsig_vs_pde} & Comparison of the computed QoI comparison between the Eulerian solver and \bolsig\\
		\hline
		Figure \ref{fig:dsmc_pde_hl_modes} & Cross-verification results for computed higher order $l$-modes using  Eulerian and DSMC solvers \\
		\hline
		Figure \ref{fig:perturb_analysis} & Relaxation times for the perturbations in higher order $l$-modes, specifically for $l>1$ with $E/n_0$ values 10Td and 20Td \\
		\hline
		Figure \ref{fig:transient_vs_steady_state} & Discrepancies for the temperature-based interpolations with tabulated steady-state data comparison to the time-dependent Boltzmann solver for oscillatory electric fields with and without Coulomb interactions\\
		\hline 
	\end{tabular}
	\caption{A summary of the presented experiments. \label{tab:summary_experiments}}    
\end{table}

\subsection{Simulation setup}
This section summarizes the simulation setup for the numerical studies presented in this paper. For simplicity, we consider background gas entirely composed of argon (Ar), where the gas density is set to be $n_0 = 3.22\times 10^{22} m^{-3}$. We assume azimuthal symmetry and isotropic scattering for all simulations presented. Under the above assumptions, the spatially homogeneous Boltzmann equation reduces to 2 dimensions in velocity space. Due to azimuthal symmetry, we use $m=0$ mode in spherical basis expansion with a specified number of $l$-modes. For $l=\{0,1\}$, we get the traditional two-term approximation~\cite{hagelaar2005solving}, but the presented methodology extends beyond the two-term approximation.
\begin{table}[!tbhp]
  \centering
	\begin{tabular}{|c|c|c|}
		\hline 
		Collision & Threshold & Post-collision\\
		\hline 
		$e + Ar \rightarrow e + Ar$ & None & $\varepsilon_{\text{post}} =\varepsilon_{\text{pre}}(1- \frac{2m_e}{m_0})$\\
		\hline
		$e + Ar \rightarrow Ar^{+} + 2e$ & $\varepsilon_{ion}$ &  $\varepsilon_{\text{post}}=0.5 \of{\varepsilon_{\text{pre}}-\varepsilon_{\text{ion}}}$\\     
		\hline
	\end{tabular}
	\caption{Electron-heavy collisions considered for the presented numerical results. \label{t:col_list}}  
\end{table}
In our experiments, we consider the following reactions 1). Elastic collisions $e + Ar \rightarrow e + Ar$ and 2). Ionization collisions $e + Ar \rightarrow Ar^{+} + 2e$ with equal energy sharing (see \Cref{t:col_list}). The steady-state solution of the system can be characterized by the effective electric field $E/n_0$ and the ionization degree $n_e/n_0$. We remark that the higher the ionization degree the more important the Coulombic interactions and thus the stronger the nonlinear effects become.  The electron distribution function can be used to compute other quantities of interest (QoIs). We focus on the mean energy ($\langle\varepsilon\rangle$ [eV]), electron mobility ($\mu_e$ $[n_0(1/m/V/s)]$), and reaction rates for process $k$ ($r_k$ $[m^3s^{-1}]$) as defined below.
\begin{align}
	\langle\varepsilon\rangle = \int_{0}^{\infty} \varepsilon^{3/2} f_0(\varepsilon) \diff{\varepsilon} \text{\ \ \ ,\ \ \ } \mu_e = -\frac{\gamma}{3(E/n_0)} \int_{0}^{\infty} \varepsilon f_1 d\varepsilon \text{\ \ \  ,\ \ \  } r_{k} = \int_{0}^{\infty} \varepsilon \sigma_k f_0 d\varepsilon \label{eq:qois}.
\end{align}

\subsection{Convergence of the Eulerian Boltzmann solver}
\label{subsec:conv_pde}
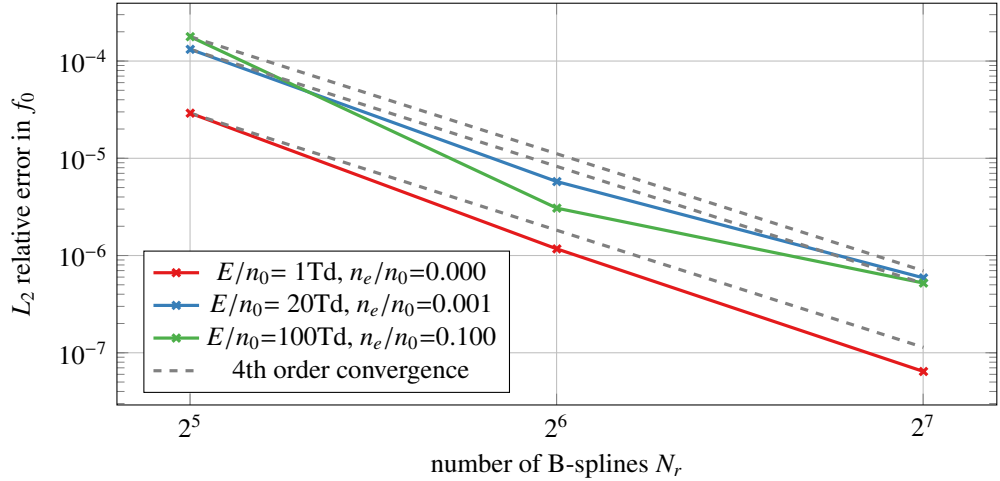
\begin{figure}[!tbhp]
	\centering
	\begin{tikzpicture}
		\begin{loglogaxis}[
			xlabel=number of B-splines $N_r$,
			ylabel=$L_2$ relative error in $f_0$, 
			log basis x={2}, width=0.8\textwidth, height=0.42\textwidth, legend pos=south west,grid=major]
			\addplot[color=a1,mark=x, very thick] coordinates {
				(32 , 0.0000291)
				(64 , 0.00000117)
				(128, 0.0000000642)
			};
			
			\addplot[color=a2,mark=x, very thick] coordinates {
				(32 , 0.000132)
				(64 , 0.00000576)
				(128, 0.000000589)
			};
			
			\addplot[color=a3,mark=x, very thick] coordinates {
				(32 , 0.000178)
				(64 , 0.00000307)
				(128, 0.000000521)
			};
			
			\addplot[color=gray, dashed, very thick] coordinates {
				(32 , 0.000178)
				(64 , 0.000178/16)
				(128, 0.000178/256)
			};
			
			\addplot[color=gray, dashed, very thick] coordinates {
				(32 , 0.000132)
				(64 , 0.000132/16)
				(128, 0.000132/256)
			};
			
			\addplot[color=gray, dashed, very thick] coordinates {
				(32 , 0.0000291)
				(64 , 0.0000291/16)
				(128, 0.0000291/256)
			};
			\legend{{$E/n_0$=   1Td, $n_e/n_0$=0.000}, {$E/n_0$= 20Td, $n_e/n_0$=0.001}, {$E/n_0$=100Td, $n_e/n_0$=0.100},  4th order convergence}
		\end{loglogaxis}
	\end{tikzpicture}
	\caption{Rate of convergence for several selected runs from \Cref{tab:self_convergence}. The above shows the rate of convergence of the relative $L_2$ error (i.e., taking $N_r$=256 as the exact solution) of the electron energy density function (i.e., $f_0$) computed with fixed two-term approximation with increasing resolution in the radial direction with cubic B-splines. As expected, the $L_2$ norm shows fourth order convergence rate for the cubic B-splines. \label{fig:convergence_rate}}
\end{figure}

We present a self-convergence study for our Eulerian solver. We use B-splines with spherical harmonics basis to represent the EDF in radial and angular directions respectively. For smooth cross-sections, the solver has exponential convergence in the angular directions. In the radial direction it has  $(p+1)^{\text{th}}$ order convergence rate (i.e., in $L_2$ norm) for $p^{\text{th}}$ order B-splines,. We use cubic B-splines (i.e., $p$=3) for all the results presented in the paper. First, we consider the traditional two-term approximation convergence with increasing refinement in the radial direction. The radial coordinate refinement can be controlled by the number of B-splines used (i.e., denoted by $N_r$).  The relative errors for the computed QoIs with varying $E/n_0$ and $n_e/n_0$ are summarized in \Cref{tab:self_convergence}. These errors are computed by taking the highest resolution run as the true solution. \Cref{fig:convergence_rate} shows the $L_2$ convergence rate of the $f_0(\vr)$ radial component (i.e., electron energy density function) with increase of radial resolution. As expected we observe fourth order convergence rate for the developed solver. \Cref{fig:ho_modes_convergence_wo_cc_terms} shows the convergence of the computed steady-state radial components (i.e., $f_l(\vr)$) with respect to the number of spherical basis functions (i.e., angular modes) used. \Cref{fig:hl_convergence_4term} shows converge of the computed higher order modes for a fixed four term approximation with increase resolution in the radial direction. For the spatially homogeneous BTE, the electric field is the key source that triggers anisotropy, while the underlying collisions counteract the triggered anisotropy (i.e., due to isotropic scattering). For the spatially coupled Boltzmann equation, there can be additional sources of anisotropy, such as boundary conditions in the physical space. Therefore, the ability to compute higher-order anisotropic correction modes can be essential for accurate solutions for a wide range of applications. 

\begin{table}[!tbhp]
	\centering
	\renewcommand{\arraystretch}{1.2}
	\begin{tabular}{|p{1cm}|p{1cm}|c|c|c|c|c|c|}
		\hline
		\multirow{2}{1cm}{\textbf{$E/n_0$ (Td)}} & \multirow{2}{1cm}{\boldmath$n_e/n_0$} & \multicolumn{6}{c|}{\textbf{relative error}}  \\
		\cline{3-8}
		& & \textbf{energy} &\textbf{mobility} & \textbf{elastic} & \textbf{ionization} & {$\mathbf{f_0(\varepsilon)}$} & {$\mathbf{f_1(\varepsilon)}$}\\
		\hline
		1	  &  0 & 	3.40E-11 & 	3.30E-11	&  5.74E-11	&   -	     &   6.42E-08	&1.99E-07\\
		5	  &  0 & 	2.01E-10 & 	8.67E-10	&  2.69E-10	& 1.05E-08 &   1.08E-06	&3.77E-06\\
		20	&  0 &  3.64E-10 & 	1.55E-09	&  6.08E-10	& 1.58E-08 &   1.78E-07	&2.50E-06\\
		100 &  0 & 	3.62E-09 & 	5.10E-09	&  7.73E-09	& 1.20E-07 &   6.51E-07	&3.32E-06\\
		\hline                  
		1   &	$10^{-3}$ &	6.69E-07 &	7.21E-07 &	9.77E-07 &	4.77E-05 &	5.82E-07 &	1.28E-06\\
		5   &	$10^{-3}$ &	9.71E-08 &	1.08E-07 &	1.40E-07 &	1.59E-06 &	1.22E-07 &	2.00E-07\\
		20  &	$10^{-3}$ &	2.02E-09 &	4.09E-09 &	4.35E-09 &	7.94E-08 &	5.89E-07 &	5.23E-07\\
		100 &	$10^{-3}$ &	6.30E-09 &	4.08E-08 &	1.33E-09 &	2.21E-08 &	1.69E-05 &  1.82E-05\\
		\hline
		1   &	$10^{-2}$ &	9.57E-06 &	1.07E-05 &	1.39E-05 &	3.36E-04 &	8.29E-06 &	1.94E-05\\
		5   &	$10^{-2}$ &	4.04E-07 &	4.32E-07 &	5.87E-07 &	9.27E-06 &	3.65E-07 &	7.97E-07\\
		20  &	$10^{-2}$ &	6.97E-08 &	6.73E-08 &	8.67E-08 &	1.76E-06 &	1.70E-07 &	2.23E-07\\
		100 &	$10^{-2}$ &	4.04E-08 &	5.47E-08 &	4.67E-08 &	1.76E-07 &	3.05E-06 &	1.67E-06\\
		\hline
		1   &	$10^{-1}$ &	8.13E-05 &	1.07E-04 &	1.17E-04 &	1.69E-03 &	7.04E-05 &	1.94E-04\\
		5   &	$10^{-1}$ &	3.34E-06 &	4.23E-06 &	4.84E-06 &	5.24E-05 &	2.89E-06 &	7.61E-06\\
		20  &	$10^{-1}$ &	7.67E-07 &	8.58E-07 &	1.04E-06 &	1.07E-05 &	6.82E-07 &	1.62E-06\\
		100 &	$10^{-1}$ &	4.04E-08 &	1.79E-08 &	2.06E-08 &	6.06E-07 &	5.21E-07 &	7.14E-07\\
		\hline
	\end{tabular}
	\caption{The table summarizes the relative errors for the QoIs computed from the steady-state solutions for different $E/n_0$ and $ne/n_0$ values with increasing refinement in the radial coordinate. The relative error is computed by $\frac{|QoI_{N_r=128} - QoI_{N_r=256}|}{|QoI_{Nr=256}|}$, where $QoI_{Nr=128}$ and $QoI_{N_r=256}$ denote QoIs computed with 128, and 256 basis functions in the radial coordinate, where relative errors for the radial components are computed by the $L^2$-norm in the energy space.  \label{tab:self_convergence}}
\end{table}

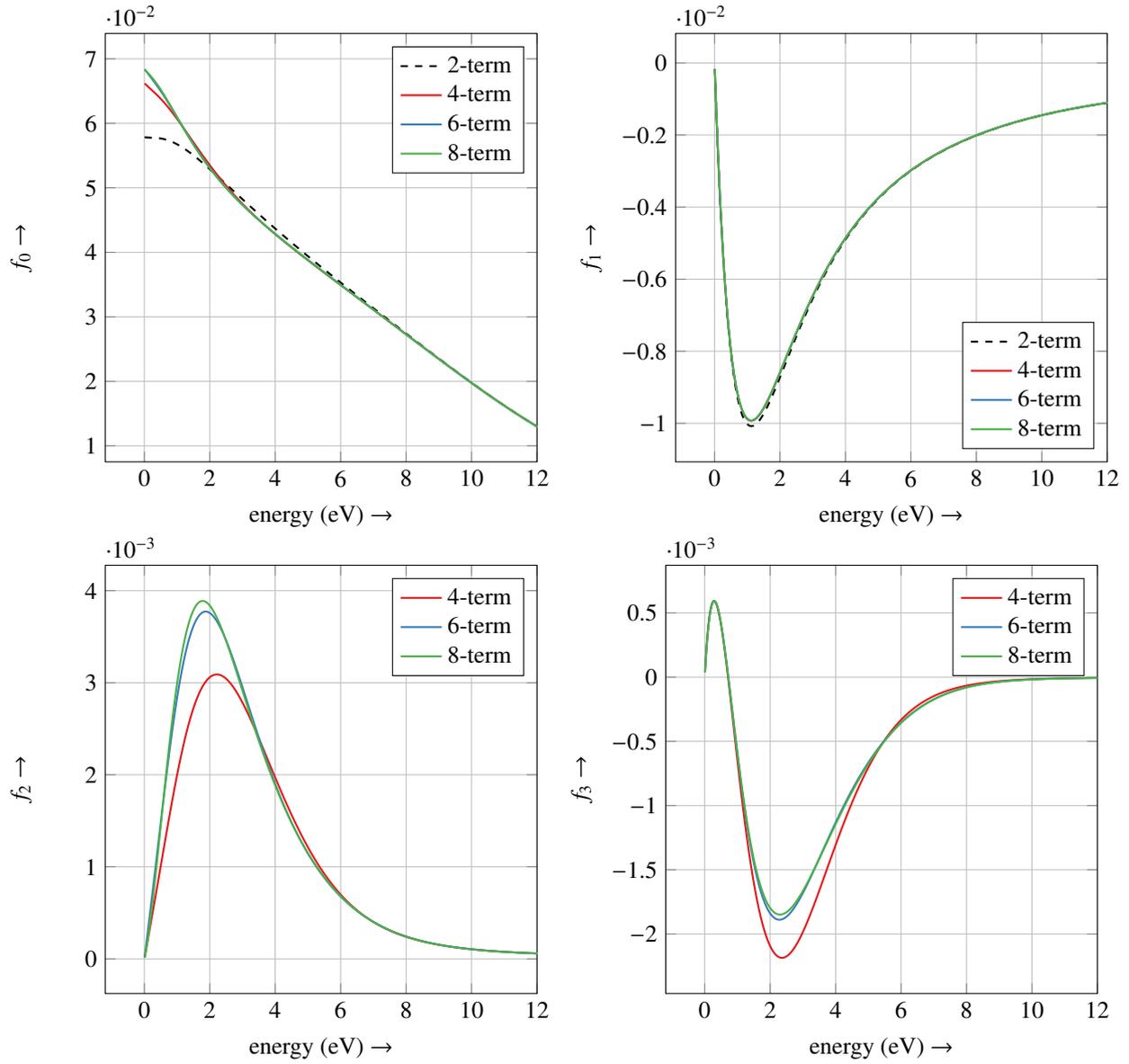
\begin{figure}[!tbhp]
  \begin{tikzpicture}
    \begin{axis}[xlabel=energy (eV) $\rightarrow$, ylabel= $f_{0}$ $\rightarrow$, grid=major,width = 0.48\textwidth,height=0.48\textwidth,xmax=12, legend pos=north east]
      \addplot[dashed,black ,thick] table[x={ev} , y ={f0} ]{pde_runs/ss_100Td_nr256_lmax1_E3.22E+03_id_0.00E+00_Tg0.00E+00.csv};
      \addplot[-,a1  ,thick] table[x={ev} , y ={f0} ]{pde_runs/ss_100Td_nr256_lmax3_E3.22E+03_id_0.00E+00_Tg0.00E+00.csv};
      \addplot[-,a2  ,thick] table[x={ev} , y ={f0} ]{pde_runs/ss_100Td_nr256_lmax5_E3.22E+03_id_0.00E+00_Tg0.00E+00.csv};
      \addplot[-,a3  ,thick] table[x={ev} , y ={f0} ]{pde_runs/ss_100Td_nr256_lmax7_E3.22E+03_id_0.00E+00_Tg0.00E+00.csv};
      \legend{2-term, 4-term, 6-term, 8-term}
    \end{axis}
  \end{tikzpicture}
  \begin{tikzpicture}
    \begin{axis}[xlabel=energy (eV) $\rightarrow$, ylabel= $f_{1}$ $\rightarrow$, grid=major,width = 0.48\textwidth,height=0.48\textwidth,xmax=12, legend pos=south east]
      \addplot[dashed,black ,thick] table[x={ev} , y ={f1} ]{pde_runs/ss_100Td_nr256_lmax1_E3.22E+03_id_0.00E+00_Tg0.00E+00.csv};
      \addplot[-,a1   ,thick] table[x={ev} , y ={f1} ]{pde_runs/ss_100Td_nr256_lmax3_E3.22E+03_id_0.00E+00_Tg0.00E+00.csv};
      \addplot[-,a2   ,thick] table[x={ev} , y ={f1} ]{pde_runs/ss_100Td_nr256_lmax5_E3.22E+03_id_0.00E+00_Tg0.00E+00.csv};
      \addplot[-,a3   ,thick] table[x={ev} , y ={f1} ]{pde_runs/ss_100Td_nr256_lmax7_E3.22E+03_id_0.00E+00_Tg0.00E+00.csv};
      \legend{2-term, 4-term, 6-term, 8-term}
    \end{axis}
  \end{tikzpicture}
  \begin{tikzpicture}
    \begin{axis}[xlabel=energy (eV) $\rightarrow$, ylabel= $f_{2}$ $\rightarrow$, grid=major,width = 0.48\textwidth,height=0.48\textwidth,xmax=12, legend pos=north east]
      \addplot[-,a1 ,thick] table[x={ev} , y ={f2} ]{pde_runs/ss_100Td_nr256_lmax3_E3.22E+03_id_0.00E+00_Tg0.00E+00.csv};
      \addplot[-,a2 ,thick] table[x={ev} , y ={f2} ]{pde_runs/ss_100Td_nr256_lmax5_E3.22E+03_id_0.00E+00_Tg0.00E+00.csv};
      \addplot[-,a3 ,thick] table[x={ev} , y ={f2} ]{pde_runs/ss_100Td_nr256_lmax7_E3.22E+03_id_0.00E+00_Tg0.00E+00.csv};
      \legend{4-term, 6-term, 8-term}
    \end{axis}
  \end{tikzpicture}
  \begin{tikzpicture}
    \begin{axis}[xlabel=energy (eV) $\rightarrow$, ylabel= $f_{3}$ $\rightarrow$, grid=major,width = 0.48\textwidth,height=0.48\textwidth,xmax=12, legend pos=north east]
      \addplot[-, a1 ,thick] table[x={ev} , y ={f3} ]{pde_runs/ss_100Td_nr256_lmax3_E3.22E+03_id_0.00E+00_Tg0.00E+00.csv};
      \addplot[-, a2 ,thick] table[x={ev} , y ={f3} ]{pde_runs/ss_100Td_nr256_lmax5_E3.22E+03_id_0.00E+00_Tg0.00E+00.csv};
      \addplot[-, a3 ,thick] table[x={ev} , y ={f3} ]{pde_runs/ss_100Td_nr256_lmax7_E3.22E+03_id_0.00E+00_Tg0.00E+00.csv};
      \legend{4-term, 6-term, 8-term}
    \end{axis}
  \end{tikzpicture}
  \caption{Convergence of steady-state EDF radial components $f_l\of{\vr}$ with respect to the number of angular terms used in the approximation. The input parameters for the above run is given by $E/N$=100Td and $n_e/n_0=0$.  	 \label{fig:ho_modes_convergence_wo_cc_terms} }
\end{figure}
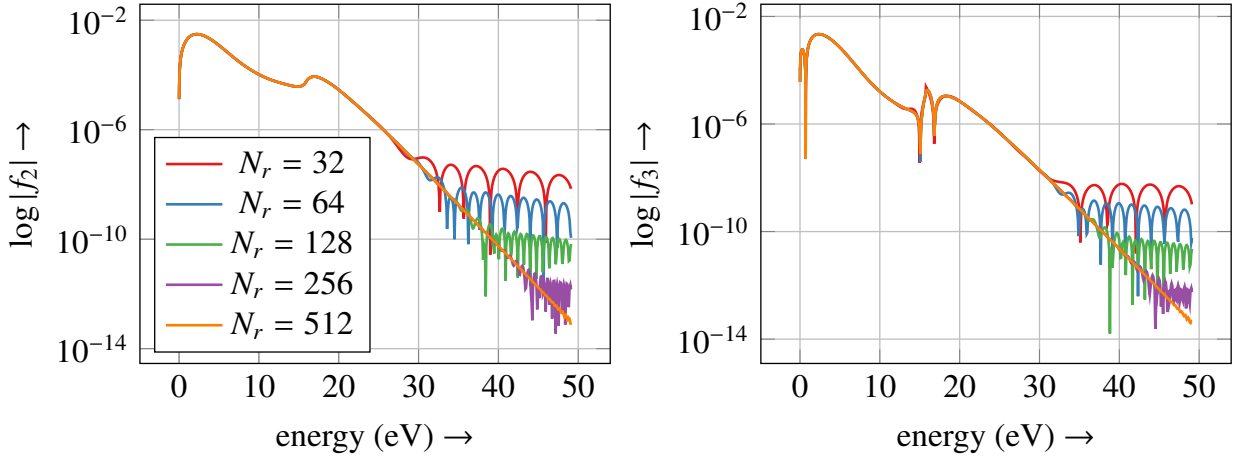
\begin{figure}[!tbhp]
	\resizebox{\textwidth}{!}{
		\begin{tikzpicture}
			\begin{semilogyaxis}[xlabel=energy (eV) $\rightarrow$, ylabel= $\log{|f_{2}|}$ $\rightarrow$, grid=major,width = 0.4\textwidth,height=0.33\textwidth, legend pos=south west]
				\addplot[-,a1 , thick] table[x={ev}, y expr=abs(\thisrow{f2})]{pde_runs/ss_100Td_nr32_lmax3_E3.22E+03_id_0.00E+00_Tg0.00E+00.csv};
				\addplot[-,a2 , thick] table[x={ev}, y expr=abs(\thisrow{f2})]{pde_runs/ss_100Td_nr64_lmax3_E3.22E+03_id_0.00E+00_Tg0.00E+00.csv};
				\addplot[-,a3 , thick] table[x={ev}, y expr=abs(\thisrow{f2})]{pde_runs/ss_100Td_nr128_lmax3_E3.22E+03_id_0.00E+00_Tg0.00E+00.csv};
				\addplot[-,a4 , thick] table[x={ev}, y expr=abs(\thisrow{f2})]{pde_runs/ss_100Td_nr256_lmax3_E3.22E+03_id_0.00E+00_Tg0.00E+00.csv};
				\addplot[-,a5 , thick] table[x={ev}, y expr=abs(\thisrow{f2})]{pde_runs/ss_100Td_nr512_lmax3_E3.22E+03_id_0.00E+00_Tg0.00E+00.csv};
				\legend{$N_r=32$, $N_r=64$, $N_r=128$, $N_r=256$, $N_r=512$}
			\end{semilogyaxis}
		\end{tikzpicture}
		\begin{tikzpicture}
			\begin{semilogyaxis}[xlabel=energy (eV) $\rightarrow$, ylabel= $\log{|f_{3}|}$ $\rightarrow$, grid=major,width = 0.4\textwidth,height=0.33\textwidth, legend pos=south west]
				\addplot[-,a1 , thick] table[x={ev}, y expr=abs(\thisrow{f3}) ]{pde_runs/ss_100Td_nr32_lmax3_E3.22E+03_id_0.00E+00_Tg0.00E+00.csv};
				\addplot[-,a2 , thick] table[x={ev}, y expr=abs(\thisrow{f3}) ]{pde_runs/ss_100Td_nr64_lmax3_E3.22E+03_id_0.00E+00_Tg0.00E+00.csv};
				\addplot[-,a3 , thick] table[x={ev}, y expr=abs(\thisrow{f3}) ]{pde_runs/ss_100Td_nr128_lmax3_E3.22E+03_id_0.00E+00_Tg0.00E+00.csv};
				\addplot[-,a4 , thick] table[x={ev}, y expr=abs(\thisrow{f3}) ]{pde_runs/ss_100Td_nr256_lmax3_E3.22E+03_id_0.00E+00_Tg0.00E+00.csv};
				\addplot[-,a5 , thick] table[x={ev}, y expr=abs(\thisrow{f3}) ]{pde_runs/ss_100Td_nr512_lmax3_E3.22E+03_id_0.00E+00_Tg0.00E+00.csv};
			\end{semilogyaxis}
		\end{tikzpicture}
	}
	\caption{Computed higher order $l$-modes (i.e., $f_2$ and $f_3$ modes) using four-term approximation with increasing resolution in the radial coordinate for input parameters given by $E/n_0$=100Td, and $n_e/n_0=0$. Increased radial resolution helps to capture the high energy tails of the computed $l$-modes. \label{fig:hl_convergence_4term} }
\end{figure}

\subsection{Performance evaluation for the Eulerian Boltzmann solver}
This section presents performance evaluation results for the developed Eulerian BTE solver. The key computational kernels for the Eulerian solver are 1). the BTE right-hand-side evaluation, 2). the Jacobian assembly, and 3). LU-factored solve of the Jacobian. We focus on the overall steady-state solve time, with overall cost breakdown for right-hand-side evaluation, Jacobian assembly, and Jacobian solve with LU factorization. The above-described operations would be the key building blocks for an evolved transient BTE solver. We conducted our experiments on Lonestar~6  at Texas Advanced Computing Center (TACC). Each node on Lonestar~6 has three NVIDIA A100 GPUs. The presented results are run on a single NVIDIA A100 GPU. \Cref{tab:ss_perfornance} presents an overall cost breakdown for a steady state BTE solve for an effective electric field of 1.8Td and with $10^{-4}$ ionization degree. The constant runtime (with increased compute throughput) for the right-hand-side and Jacobian assembly indicates the GPU is not fully saturated for the specified input sizes presented in \Cref{tab:ss_perfornance}. The overall solve cost scales as $\mathcal{O}(N_v^3)$, hence leading to an increase in the overall solve time with increased $\vect{v}$-space resolution. 

\begin{table}[!tbhp]
	\centering
	\resizebox{\textwidth}{!}{
	\begin{tabular}{|c|c|c|c|c|c|c|c|c|c|}
	\hline
	$\bm{N_r \times N_l}$ & \textbf{solve} (s) & \textbf{RHS/call (s)} & \textbf{Flops/s (RHS)} & \textbf{$\vect{J}$/call (s)} & \textbf{Flops/s ($\vect{J}$)} & \textbf{$\vect{J}$-LU / call} (s) & \textbf{Flops/s ($\vect{J}$-LU)}\\
	\hline
	64 $\times$ 2  & 0.79063 & 0.001436 & 3.05E+09 & 0.0023765 & 7.03E+09 & 0.002713 & 7.67E+08 \\
	\hline
    128 $\times$ 2 & 0.90317 & 0.001382 & 2.48E+10 & 0.0020234 & 6.62E+10 & 0.008581 & 1.95E+09 \\
    \hline 
    256 $\times$ 2 & 3.15850 & 0.001414 & 1.92E+11 & 0.0037712 & 2.84E+11 & 0.049841 & 2.69E+09 \\
    \hline
	\end{tabular}}
\caption{Summary of the overall solve cost for the steady-state Eulerian BTE solver. \label{tab:ss_perfornance}}
\end{table}

\subsection{Performance evaluation for the DSMC Boltzmann solver}
We now present performance evaluation results for the DSMC BTE solver. Unlike the Eulerian solver, the DSMC solver does not support direct steady-state computations. Instead, a system is initialized using a Maxwellian distribution near the expected steady-state temperature, and time steps are taken until the system is converged. For the above, the convergence is determined by the relative error of macroscopic quantities (i.e., temperature, reaction rates, and mobility). We conducted our experiments on Lassen at Lawrence Livermore National Laboratory (LLNL). Each node on Lassen has four NVIDIA Volta V100 GPUs; the presented results are run on a single GPU. \Cref{tab:dsmc_performance} presents the runtime to steady state for four different electric field values. Notably, the run time to steady state is highly dependent on the magnitude of the reduced electric field. This behavior is due to the ability of a higher electric field to push a system to a steady state in less time. Additionally, the lower temperature systems that result from lower electric fields often require more particles and more samples to accurately resolve the ionization reaction rate coefficient. This increased resolution further leads to greater computational costs.

\begin{table}[!tbhp]
	\centering
	\begin{tabular}{|c|c|c|c|}
	\hline
	$\frac{E}{n_0}$ & $\frac{n_e}{n_0}$ & \textbf{number of initial particles} & \textbf{solve} (s) \\
	\hline
	1Td & $10^{-3}$ & 1E+06 & 15405 \\
	\hline
    5Td & $10^{-3}$ & 1E+05 & 1729 \\
    \hline 
    20Td & $10^{-3}$ & 1E+05 & 127  \\
    \hline 
    100Td & $10^{-3}$ & 1E+05 & 76  \\
    \hline
	\end{tabular}
\caption{Summary of the overall time to solution for the DSMC BTE solver for different electric field values. \label{tab:dsmc_performance}}
\end{table}

\subsection{Convergence of the DSMC solver}
\label{subsec:conv_dsmc}
For the DSMC code, convergence testing is limited to examining one parameter, the number of samples used. For self-convergence, we consider the exact solution to be the solution given by 1 billion ($10^{9}$) samples. These results were averaged over 100 separate samples of a given sample size to ensure reproducibility and accuracy. Table~\ref{tab:dsmc_self_conv2} summarizes the convergence results for the DSMC solver. 
\begin{table}[!tbhp]
  \begin{minipage}{0.48\textwidth}
    \centering
    \resizebox{\textwidth}{!}{
    \begin{tabular}{|c|c|c|c|c|}
      \hline
      \textbf{\# of samples} & \textbf{temperature} & \textbf{elastic} & \textbf{ionization} & \textbf{mobility} \\
      \hline
      $10^3$ & 1.95E-02 & 4.06E-01 & 1.37E+00 & 2.02E-01  \\
      $10^4$ & 6.08E-03 & 1.08E-01 & 1.44E+00 & 5.29E-02	\\
      $10^5$ & 2.03E-03 & 3.23E-02 & 4.75E-01 & 1.82E-02	\\
      $10^6$ & 6.34E-04 & 1.07E-02 & 1.40E-01 & 5.38E-03	\\
      $10^7$ & 1.89E-04 & 4.25E-03 & 6.82E-02 & 2.78E-03  \\
      $10^8$ & 2.94E-05 & 1.35E-05 & 2.74E-03 & 1.16E-03  \\
      \hline
    \end{tabular}}
  \end{minipage}
  \begin{minipage}{0.48\textwidth}
    \resizebox{\textwidth}{!}{
    \begin{tabular}{|c|c|c|c|c|}
      \hline
      \textbf{\# of samples} & \textbf{temperature} & \textbf{elastic} & \textbf{ionization} & \textbf{mobility} \\
      \hline
      $10^6$ & 6.97E-04 & 1.52E-02 & 1.00E+00 & 3.60E-02	\\
      $10^7$ & 7.61E-04 & 2.84E-03 & 1.66E-01 & 2.63E-02	\\
      $10^8$ & 6.91E-04 & 2.62E-03 & 9.44E-02 & 2.58E-02	\\
      $10^9$ & 6.24E-04 & 1.51E-03 & 2.97E-02 & 1.45E-02	\\
      $10^{10}$ & 1.77E-04 & 1.03E-03 & 5.83E-03 & 8.91E-03  \\
      $10^{11}$ & 2.03E-05 & 3.13E-05 & 7.38E-04 & 2.47E-04  \\
      \hline
    \end{tabular}}
  \end{minipage}
  \caption{Self convergence of the DSMC solver with respect to the number of sampled used for QoI computation. In the above, the left and the right table shows convergence results for $(\frac{E}{n_0} = 100 \mathrm{Td}, \frac{n_e}{n_0} = 10^{-1})$ and $(\frac{E}{n_0} = 5\mathrm{Td}, \frac{n_e}{n_0} = 10^{-2})$ respectively. \label{tab:dsmc_self_conv2}}
\end{table}

\begin{figure}[!tbhp]
  \begin{center}
    \begin{tikzpicture}
      \begin{loglogaxis}[xlabel= $E/n_0$(Td) $\rightarrow$, ylabel= steady-state energy (eV) $\rightarrow$, grid=major,width = 0.48\textwidth,height=0.48\textwidth, legend pos=south east, xmin=1e-3]
        \addplot[,a1 ,mark=*, only marks, mark size=1pt] table[x={E/N(Td)} , y ={energy} ]{pde_runs/pde_vs_bolsig_with_coulomb_collision_nr128_id0.00E+00.csv};
        \addplot[,a2 ,mark=x, only marks, mark size=2pt] table[x={E/N(Td)} , y ={bolsig_energy} ]{pde_runs/pde_vs_bolsig_with_coulomb_collision_nr128_id0.00E+00.csv};
        
        \addplot[,a6 ,mark=*, only marks, mark size=1pt] table[x={E/N(Td)} , y ={energy} ]{pde_runs/pde_vs_bolsig_with_coulomb_collision_nr64_id1.00E-03.csv};
        \addplot[,a7 ,mark=x, only marks, mark size=2pt] table[x={E/N(Td)} , y ={bolsig_energy} ]{pde_runs/pde_vs_bolsig_with_coulomb_collision_nr64_id1.00E-03.csv};
        
        \addplot[,a4 ,mark=*, only marks, mark size=1pt] table[x={E/N(Td)} , y ={energy} ]{pde_runs/pde_vs_bolsig_with_coulomb_collision_nr64_id1.00E-02.csv};
        \addplot[,a5 ,mark=x, only marks, mark size=2pt] table[x={E/N(Td)} , y ={bolsig_energy} ]{pde_runs/pde_vs_bolsig_with_coulomb_collision_nr64_id1.00E-02.csv};
        
        \addplot[,a2 ,mark=*, only marks, mark size=1pt] table[x={E/N(Td)} , y ={energy} ]{pde_runs/pde_vs_bolsig_with_coulomb_collision_nr64_id1.00E-01.csv};
        \addplot[,a3 ,mark=x, only marks, mark size=2pt] table[x={E/N(Td)} , y ={bolsig_energy} ]{pde_runs/pde_vs_bolsig_with_coulomb_collision_nr64_id1.00E-01.csv};
      \end{loglogaxis}
    \end{tikzpicture}
    \begin{tikzpicture}
      \begin{loglogaxis}[xlabel= $E/n_0$(Td) $\rightarrow$, ylabel= mobility $(N (1/m/V/s))~\rightarrow$, grid=major,width = 0.48\textwidth,height=0.48\textwidth, legend pos=south east, xmin=1e-3]
        \addplot[,a1 ,mark=*, only marks, mark size=1pt] table[x={E/N(Td)} , y ={mobility} ]{pde_runs/pde_vs_bolsig_with_coulomb_collision_nr128_id0.00E+00.csv};
        \addplot[,a2 ,mark=x, only marks, mark size=2pt] table[x={E/N(Td)} , y ={bolsig_mobility} ]{pde_runs/pde_vs_bolsig_with_coulomb_collision_nr128_id0.00E+00.csv};
        
        \addplot[,a6 ,mark=*, only marks, mark size=1pt] table[x={E/N(Td)} , y ={mobility} ]{pde_runs/pde_vs_bolsig_with_coulomb_collision_nr64_id1.00E-03.csv};
        \addplot[,a7 ,mark=x, only marks, mark size=2pt] table[x={E/N(Td)} , y ={bolsig_mobility} ]{pde_runs/pde_vs_bolsig_with_coulomb_collision_nr64_id1.00E-03.csv};
        
        \addplot[,a4 ,mark=*, only marks, mark size=1pt] table[x={E/N(Td)} , y ={mobility} ]{pde_runs/pde_vs_bolsig_with_coulomb_collision_nr64_id1.00E-02.csv};
        \addplot[,a5 ,mark=x, only marks, mark size=2pt] table[x={E/N(Td)} , y ={bolsig_mobility} ]{pde_runs/pde_vs_bolsig_with_coulomb_collision_nr64_id1.00E-02.csv};
        
        \addplot[,a2 ,mark=*, only marks, mark size=1pt] table[x={E/N(Td)} , y ={mobility} ]{pde_runs/pde_vs_bolsig_with_coulomb_collision_nr64_id1.00E-01.csv};
        \addplot[,a3 ,mark=x, only marks, mark size=2pt] table[x={E/N(Td)} , y ={bolsig_mobility} ]{pde_runs/pde_vs_bolsig_with_coulomb_collision_nr64_id1.00E-01.csv};
      \end{loglogaxis}
    \end{tikzpicture}
    \begin{tikzpicture}
      \begin{loglogaxis}[xlabel= $E/n_0$(Td) $\rightarrow$, ylabel= elastic collisions rate $(m^3s^{-1})~\rightarrow$, grid=major,width = 0.48\textwidth,height=0.48\textwidth, legend pos=south east, legend columns=1, legend style={font=\footnotesize}, xmin=1e-3]
        \addplot[,a1 ,mark=*, only marks, mark size=1pt] table[x={E/N(Td)} , y ={g0} ]{pde_runs/pde_vs_bolsig_with_coulomb_collision_nr128_id0.00E+00.csv};
        \addplot[,a2 ,mark=x, only marks, mark size=2pt] table[x={E/N(Td)} , y ={bolsig_g0} ]{pde_runs/pde_vs_bolsig_with_coulomb_collision_nr128_id0.00E+00.csv};
        
        \addplot[,a6 ,mark=*, only marks, mark size=1pt] table[x={E/N(Td)} , y ={g0} ]{pde_runs/pde_vs_bolsig_with_coulomb_collision_nr64_id1.00E-03.csv};
        \addplot[,a7 ,mark=x, only marks, mark size=2pt] table[x={E/N(Td)} , y ={bolsig_g0} ]{pde_runs/pde_vs_bolsig_with_coulomb_collision_nr64_id1.00E-03.csv};
        
        \addplot[,a4 ,mark=*, only marks, mark size=1pt] table[x={E/N(Td)} , y ={g0} ]{pde_runs/pde_vs_bolsig_with_coulomb_collision_nr64_id1.00E-02.csv};
        \addplot[,a5 ,mark=x, only marks, mark size=2pt] table[x={E/N(Td)} , y ={bolsig_g0} ]{pde_runs/pde_vs_bolsig_with_coulomb_collision_nr64_id1.00E-02.csv};
        
        \addplot[,a2 ,mark=*, only marks, mark size=1pt] table[x={E/N(Td)} , y ={g0} ]{pde_runs/pde_vs_bolsig_with_coulomb_collision_nr64_id1.00E-01.csv};
        \addplot[,a3 ,mark=x, only marks, mark size=2pt] table[x={E/N(Td)} , y ={bolsig_g0} ]{pde_runs/pde_vs_bolsig_with_coulomb_collision_nr64_id1.00E-01.csv};
        \legend{(A) $n_e/n_0=0$, (B) $n_e/n_0=0$, (A) $n_e/n_0=10^{-3}$, (B) $n_e/n_0=10^{-3}$, (A) $n_e/n_0=10^{-2}$, (B) $n_e/n_0=10^{-2}$, (A) $n_e/n_0=10^{-1}$, (B) $n_e/n_0=10^{-1}$ }
      \end{loglogaxis}
    \end{tikzpicture}
    \begin{tikzpicture}
      \begin{loglogaxis}[xlabel= $E/n_0$(Td) $\rightarrow$, ylabel= ionization collisions rate $(m^3s^{-1})~\rightarrow$, grid=major,width = 0.48\textwidth,height=0.48\textwidth, legend pos=south east,xmin=1e-3]
        \addplot[,a1 ,mark=*, only marks, mark size=1pt] table[x={E/N(Td)} , y ={g2} ]{pde_runs/pde_vs_bolsig_with_coulomb_collision_nr128_id0.00E+00.csv};
        \addplot[,a2 ,mark=x, only marks, mark size=1pt] table[x={E/N(Td)} , y ={bolsig_g2} ]{pde_runs/pde_vs_bolsig_with_coulomb_collision_nr128_id0.00E+00.csv};
        
        \addplot[,a6 ,mark=*, only marks, mark size=1pt] table[x={E/N(Td)} , y ={g2} ]{pde_runs/pde_vs_bolsig_with_coulomb_collision_nr64_id1.00E-03.csv};
        \addplot[,a7 ,mark=x, only marks, mark size=1pt] table[x={E/N(Td)} , y ={bolsig_g2} ]{pde_runs/pde_vs_bolsig_with_coulomb_collision_nr64_id1.00E-03.csv};
        
        \addplot[,a4 ,mark=*, only marks, mark size=1pt] table[x={E/N(Td)} , y ={g2} ]{pde_runs/pde_vs_bolsig_with_coulomb_collision_nr64_id1.00E-02.csv};
        \addplot[,a5 ,mark=x, only marks, mark size=1pt] table[x={E/N(Td)} , y ={bolsig_g2} ]{pde_runs/pde_vs_bolsig_with_coulomb_collision_nr64_id1.00E-02.csv};
        
        \addplot[,a2 ,mark=*, only marks, mark size=1pt] table[x={E/N(Td)} , y ={g2} ]{pde_runs/pde_vs_bolsig_with_coulomb_collision_nr64_id1.00E-01.csv};
        \addplot[,a3 ,mark=x, only marks, mark size=1pt] table[x={E/N(Td)} , y ={bolsig_g2} ]{pde_runs/pde_vs_bolsig_with_coulomb_collision_nr64_id1.00E-01.csv};
      \end{loglogaxis}
    \end{tikzpicture}
  \end{center}
	\caption{Computed QoIs from the developed Eulerian steady-state solver (i.e., denoted by A and the \textbullet\ marker) compared against the \bolsig~solver (i.e., denoted by B and the $\times$ marker) for varying $E/n_0$ and $n_e/n_0$ values.  \label{fig:bolsig_vs_pde}}
\end{figure}
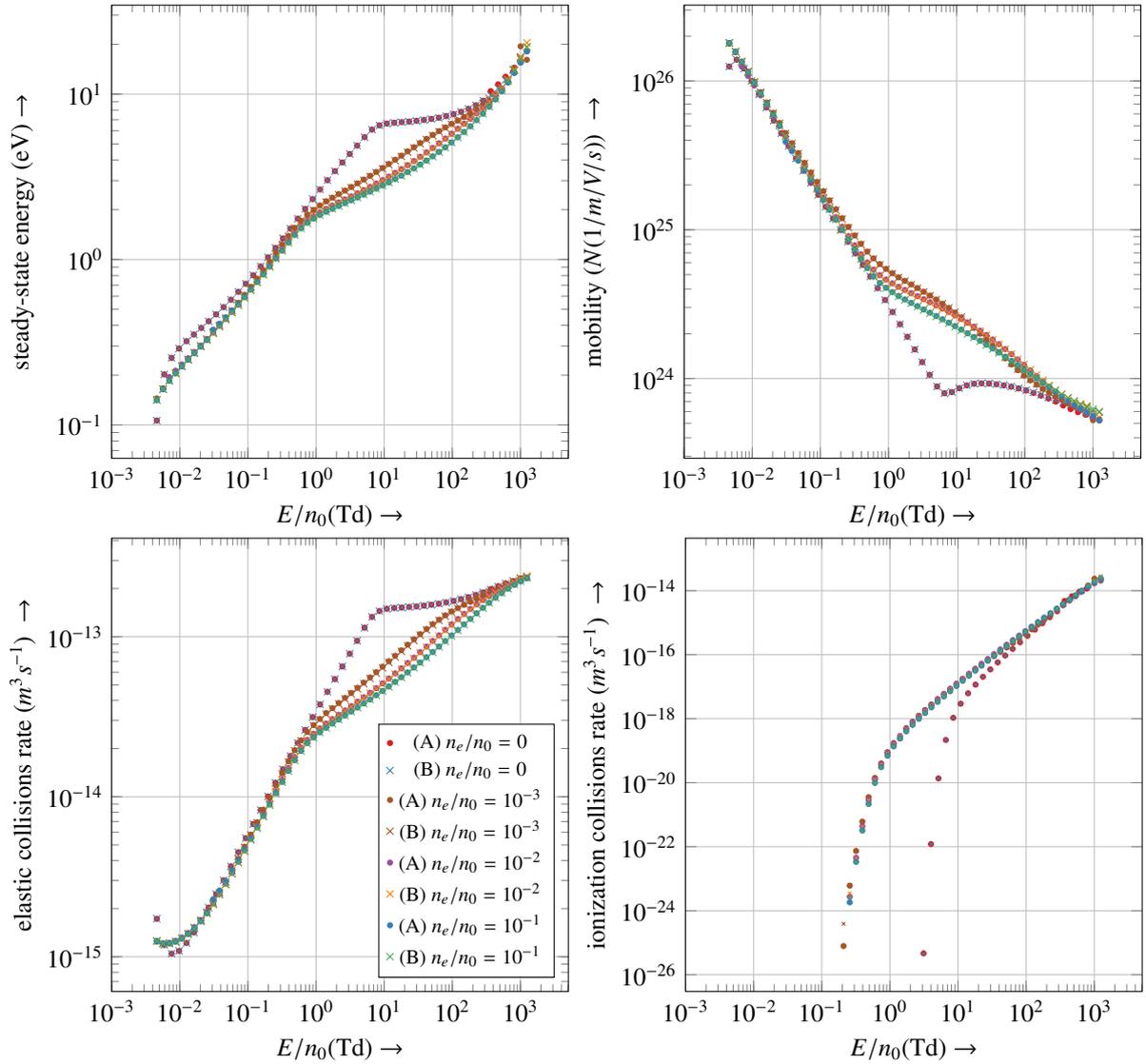

\begin{table}[!tbhp]
	\centering
	\resizebox{0.8\textwidth}{!}{
		\renewcommand{\arraystretch}{1.2}
		\begin{tabular}{|p{1cm}|p{1cm}|c|c|c|c|c|c|}
			\hline
			\multirow{2}{1cm}{\textbf{$E/n_0$ (Td)}} & \multirow{2}{1cm}{\boldmath$n_e/n_0$} & \multicolumn{3}{c|}{\textbf{rel. error (Eulerian vs. \bolsig)}} & \multicolumn{3}{c|}{\textbf{rel. error (DSMC vs. \bolsig)}} \\
			\cline{3-8}
			& & \textbf{elastic} & \textbf{ionization} & \textbf{mobility} & \textbf{elastic} & \textbf{ionization} & \textbf{mobility}\\
			\hline
			1        & 0.0 & 7.11E-05 &    --	    & 4.10E-05	& 7.79E-03	& --  & 4.80E-03 \\
			5        & 0.0 & 4.12E-04 & 3.74E-04	& 2.92E-06	& 7.77E-03	& 1.95E-03	& 1.87E-03   \\
			20       & 0.0 & 5.76E-04 & 3.49E-03	& 2.98E-04	& 8.33E-03	& 4.77E-03	& 3.92E-03  \\
			100      & 0.0 & 1.35E-03 & 2.74E-03	& 4.85E-03	& 1.14E-02	& 8.31E-03	& 5.05E-03   \\
			\hline
			1        & $10^{-3}$ & 9.40E-03	& 4.40E-03	& 7.99E-03	& 6.23E-03	& 7.61E-03	& 3.17E-03  \\
			5        & $10^{-3}$ & 1.52E-02	& 4.12E-03	& 1.83E-04	& 5.90E-03	& 6.59E-03	& 5.89E-04  \\
			20       & $10^{-3}$ & 1.69E-02	& 1.46E-02	& 1.13E-02	& 8.17E-03	& 1.09E-02	& 4.73E-03   \\
			100      & $10^{-3}$ & 9.11E-03	& 1.87E-02	& 1.65E-02	& 1.50E-02	& 2.42E-02	& 1.75E-02   \\
			\hline
			1        & $10^{-2}$ & 4.88E-03	& 1.20E-02	& 9.94E-03	& 8.66E-03  & 1.38E-02  & 4.98E-03   \\
			5        & $10^{-2}$ & 7.85E-03	& 8.73E-03	& 9.56E-03	& 4.82E-03  & 4.66-03  & 4.23E-03    \\
			20       & $10^{-2}$ & 1.13E-02	& 4.57E-03	& 6.19E-03	& 5.36E-03  & 5.72E-03  & 2.10E-03   \\
			100      & $10^{-2}$ & 1.19E-02	& 9.93E-03	& 7.76E-03	& 1.04E-02  & 2.05E-02  & 1.38E-02  \\
			\hline
			1        & $10^{-1}$ & 1.72E-03	& 8.17E-03	& 5.31E-03	& 5.80E-03  & 3.22E-02  & 6.75E-03  \\
			5        & $10^{-1}$ & 2.42E-03	& 7.80E-03	& 7.74E-03	& 3.41E-03  & 3.03E-04  & 8.80E-03   \\
			20       & $10^{-1}$ & 3.84E-03	& 7.97E-03	& 8.41E-03	& 3.30E-03  & 1.08E-02  & 9.56E-03   \\
			100      & $10^{-1}$ & 4.76E-03	& 9.80E-04	& 1.89E-03	& 5.49E-03  & 5.59E-03  & 2.64E-03   \\
			\hline
	\end{tabular}}
	\caption{The table summaries the relative errors for computed QoIs between the presented work compared to the state-of-the-art \bolsig~code. 
		\label{tab:errors_with_bolsig}}
\end{table}

\subsection{Comparison with \bolsig}
\label{subsec:verification}
Now we describe additional verification tests in which we compare our Eulerian and DSCM solvers with a state-of-the-art Eulerian Boltzmann code, \bolsig~\cite{hagelaar2005solving,hagelaar2015coulomb}. \bolsig uses a special finite-difference scheme in the radial direction and a fixed two-term expansion in the azimuthal direction. \Cref{tab:errors_with_bolsig} summarizes the relative errors for the computed QoIs with Eulerian and DSMC approaches, assuming the \bolsig~solution as the ground truth for varying $E/n_0$ and $n_e/n_0$ values. Figure~\ref{fig:bolsig_vs_pde} shows computed QoIs with varying parameter configurations (i.e., $E/n_0$ and $n_e/n_0$) corresponding \bolsig~results. For all conducted runs, the relative errors for the computed QoIs are less than 2\% between the proposed approaches and \bolsig~code. Several factors, such as boundary treatment, tabulated cross-section interpolation, and differences in numerical schemes, can contribute to the errors we observe between \bolsig~and developed Eulerian and Lagrangian solvers. The observed relative errors between codes are well within the uncertainty of cross-sections. Hence, for all practical purposes, the developed solvers produce identical results to \bolsig.

\subsection{Cross-verification for higher l-modes} 
\label{subsec:crossverify_1}
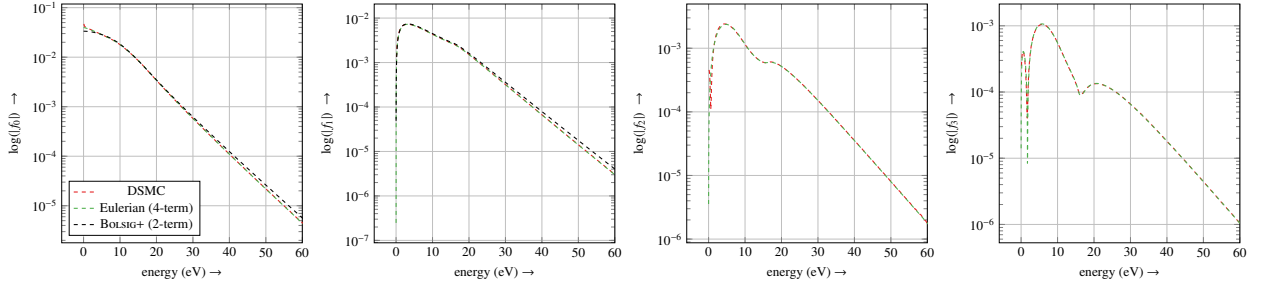
\begin{figure}[!tbhp]
	\resizebox{\textwidth}{!}{
		\begin{tikzpicture}
			\begin{semilogyaxis}[xlabel= energy (eV) $\rightarrow$, ylabel= $\log(|f_0|)~\rightarrow$, grid=major,width = 0.48\textwidth,height=0.48\textwidth, legend pos=south west, xmax=60]
				\addplot[dashed, a1, thick ] table[x={ev}, y expr=abs(\thisrow{f0}) ]{pde_runs/dsmc_500Td_0_75B_final.csv};
				\addplot[dashed, a3, thick ] table[x={ev}, y expr=abs(\thisrow{f0}) ]{pde_runs/ss_500Td_nr512_lmax5_E1.61E+04_id_0.00E+00_Tg0.00E+00.csv};
				\addplot[dashed, black , thick] table[x={ev}, y expr=abs(\thisrow{bolsig_f0}) ]{pde_runs/ss_500Td_nr512_lmax5_E1.61E+04_id_0.00E+00_Tg0.00E+00.csv};
				\legend{DSMC, Eulerian (4-term), \bolsig~(2-term)}
			\end{semilogyaxis}
		\end{tikzpicture}
		\begin{tikzpicture}
			\begin{semilogyaxis}[xlabel= energy (eV) $\rightarrow$, ylabel= $\log(|f_1|)~\rightarrow$, grid=major,width = 0.48\textwidth,height=0.48\textwidth, legend pos=south east,xmax=60]
				\addplot[dashed,a1 , thick] table[x={ev}, y expr=abs(\thisrow{f1}) ]{pde_runs/dsmc_500Td_0_75B_final.csv};
				\addplot[dashed,a3 , thick] table[x={ev}, y expr=abs(\thisrow{f1}) ]{pde_runs/ss_500Td_nr512_lmax5_E1.61E+04_id_0.00E+00_Tg0.00E+00.csv};
				\addplot[dashed, black , thick] table[x={ev}, y expr=abs(\thisrow{bolsig_f1}) ]{pde_runs/ss_500Td_nr512_lmax5_E1.61E+04_id_0.00E+00_Tg0.00E+00.csv};
			\end{semilogyaxis}
		\end{tikzpicture}
		
		\begin{tikzpicture}
			\begin{semilogyaxis}[xlabel= energy (eV) $\rightarrow$, ylabel= $\log(|f_2|)~\rightarrow$, grid=major,width = 0.48\textwidth,height=0.48\textwidth, legend pos=south east,xmax=60]
				\addplot[dashed,a1 , thick] table[x={ev}, y expr=abs(\thisrow{f2}) ]{pde_runs/dsmc_500Td_0_75B_final.csv};
				\addplot[dashed,a3 , thick] table[x={ev}, y expr=abs(\thisrow{f2}) ]{pde_runs/ss_500Td_nr512_lmax5_E1.61E+04_id_0.00E+00_Tg0.00E+00.csv};
			\end{semilogyaxis}
		\end{tikzpicture}
		\begin{tikzpicture}
			\begin{semilogyaxis}[xlabel= energy (eV) $\rightarrow$, ylabel= $\log(|f_3|)~\rightarrow$, grid=major,width = 0.48\textwidth,height=0.48\textwidth, legend pos=south east,xmax=60]
				\addplot[dashed,a1 , thick] table[x={ev}, y expr=abs(\thisrow{f3}) ]{pde_runs/dsmc_500Td_0_75B_final.csv};
				\addplot[dashed,a3 , thick] table[x={ev}, y expr=abs(\thisrow{f3}) ]{pde_runs/ss_500Td_nr512_lmax5_E1.61E+04_id_0.00E+00_Tg0.00E+00.csv};
			\end{semilogyaxis}
		\end{tikzpicture}}
		\caption{In the above simulation, we compare the computed $l$-modes between the DSMC, proposed Eulerian approach and the~\bolsig code for the case with $E/n_0$=500Td and $n_e/n_0=0$. Note that, \bolsig~code uses the two-term expansion, the proposed Eulerian approach uses multi-term expansion, and DSMC code does not inherit a term-expansion. \label{fig:dsmc_pde_hl_modes}}
\end{figure}

In this section, we present cross-verification results between our Eulerian and DSMC solvers. As previously discussed, the \bolsig~solver uses the two-term approximation that ignores the higher-order anisotropic correction terms. Our Eulerian model can capture an arbitrary number of anisotropic correction modes; here, we consider the first four modes.
We compare our Eulerian solver to our DSMC solver, which inherently has no term approximation. To study these higher order modes, we solve the steady state problem as in \Cref{subsec:ss_and_transient_sol}. Figure \ref{fig:dsmc_pde_hl_modes} shows perfect agreement between computed higher order modes between DSMC and Eulerian BTE solvers. 

\subsection{Perturbations in higher l-modes}
\label{subsec:perturbations}
\begin{figure}[!tbhp]
	\centering
  \begin{subfigure}{\textwidth}
	\resizebox{\textwidth}{!}{
		\begin{tikzpicture}
			\begin{axis}[xlabel= time (s) $\rightarrow$, ylabel= energy (eV) $\rightarrow$, grid=major,width = 0.48\textwidth, height=0.48\textwidth, legend pos=south east, yticklabel style={/pgf/number format/fixed, /pgf/number format/precision=4}]
				\addplot[-,a1  ,thick] table[x={time}, y ={energy} ]{pde_runs/pm_nr256_lmax3_E3.22E+02_id_0.00E+00_Tg0.00E+00_perturb_l_2_dt1.00E-09.csv};
				\addplot[-,a2  ,thick] table[x={time}, y ={energy} ]{pde_runs/pm_nr256_lmax3_E3.22E+02_id_0.00E+00_Tg0.00E+00_perturb_l_3_dt1.00E-09.csv};
				\legend{$l=2$ , $l=3$}
			\end{axis}
		\end{tikzpicture}
		\begin{tikzpicture}
			\begin{axis}[xlabel= time (s) $\rightarrow$, ylabel= elastic rate ($m^3s^{-1}$) $\rightarrow$, grid=major,width=0.48\textwidth,height=0.48\textwidth, yticklabel style={/pgf/number format/fixed, /pgf/number format/precision=6}]
        \addplot[-,a1  ,thick] table[x={time}, y ={g0} ]{pde_runs/pm_nr256_lmax3_E3.22E+02_id_0.00E+00_Tg0.00E+00_perturb_l_2_dt1.00E-09.csv};
				\addplot[-,a2  ,thick] table[x={time}, y ={g0} ]{pde_runs/pm_nr256_lmax3_E3.22E+02_id_0.00E+00_Tg0.00E+00_perturb_l_3_dt1.00E-09.csv};
			\end{axis}
		\end{tikzpicture}
		\begin{tikzpicture}
			\begin{axis}[xlabel= time (s) $\rightarrow$, ylabel= ionization rate ($m^3s^{-1}$ )$\rightarrow$, grid=major,width=0.48\textwidth,height=0.48\textwidth, yticklabel style={/pgf/number format/fixed, /pgf/number format/precision=6}]
				\addplot[-,a1  ,thick] table[x={time}, y ={g2} ]{pde_runs/pm_nr256_lmax3_E3.22E+02_id_0.00E+00_Tg0.00E+00_perturb_l_2_dt1.00E-09.csv};
				\addplot[-,a2  ,thick] table[x={time}, y ={g2} ]{pde_runs/pm_nr256_lmax3_E3.22E+02_id_0.00E+00_Tg0.00E+00_perturb_l_3_dt1.00E-09.csv};
			\end{axis}
		\end{tikzpicture}
	}
  \end{subfigure}
  \begin{subfigure}{\textwidth}
    \resizebox{\textwidth}{!}{
      \begin{tikzpicture}
        \begin{axis}[xlabel= time (s) $\rightarrow$, ylabel= energy (eV) $\rightarrow$, grid=major,width = 0.48\textwidth, height=0.48\textwidth, legend pos=south east, yticklabel style={/pgf/number format/fixed, /pgf/number format/precision=4}]
          \addplot[-,a1  ,thick] table[x={time}, y ={energy} ]{pde_runs/pm_nr256_lmax3_E6.44E+02_id_0.00E+00_Tg0.00E+00_perturb_l_2_dt1.00E-09.csv};
          \addplot[-,a2  ,thick] table[x={time}, y ={energy} ]{pde_runs/pm_nr256_lmax3_E6.44E+02_id_0.00E+00_Tg0.00E+00_perturb_l_3_dt1.00E-09.csv};
          \legend{$l=2$ , $l=3$}
        \end{axis}
      \end{tikzpicture}
      \begin{tikzpicture}
        \begin{axis}[xlabel= time (s) $\rightarrow$, ylabel= elastic rate ($m^3s^{-1}$) $\rightarrow$, grid=major,width=0.48\textwidth,height=0.48\textwidth, yticklabel style={/pgf/number format/fixed, /pgf/number format/precision=6}]
          \addplot[-,a1  ,thick] table[x={time}, y ={g0} ]{pde_runs/pm_nr256_lmax3_E6.44E+02_id_0.00E+00_Tg0.00E+00_perturb_l_2_dt1.00E-09.csv};
          \addplot[-,a2  ,thick] table[x={time}, y ={g0} ]{pde_runs/pm_nr256_lmax3_E6.44E+02_id_0.00E+00_Tg0.00E+00_perturb_l_3_dt1.00E-09.csv};
        \end{axis}
      \end{tikzpicture}
      \begin{tikzpicture}
        \begin{axis}[xlabel= time (s) $\rightarrow$, ylabel= ionization rate ($m^3s^{-1}$ )$\rightarrow$, grid=major,width=0.48\textwidth,height=0.48\textwidth, yticklabel style={/pgf/number format/fixed, /pgf/number format/precision=6}]
          \addplot[-,a1  ,thick] table[x={time}, y ={g2} ]{pde_runs/pm_nr256_lmax3_E6.44E+02_id_0.00E+00_Tg0.00E+00_perturb_l_2_dt1.00E-09.csv};
          \addplot[-,a2  ,thick] table[x={time}, y ={g2} ]{pde_runs/pm_nr256_lmax3_E6.44E+02_id_0.00E+00_Tg0.00E+00_perturb_l_3_dt1.00E-09.csv};
        \end{axis}
      \end{tikzpicture}
    }
    \end{subfigure}
	\caption{Perturbation study conducted on the higher-order $l$-modes for a case with $E/n_0$=10Td (in the top) and $E/n_0$=20Td (in the bottom) with $n_e / n_0$ = 0. \label{fig:perturb_analysis}}
\end{figure}
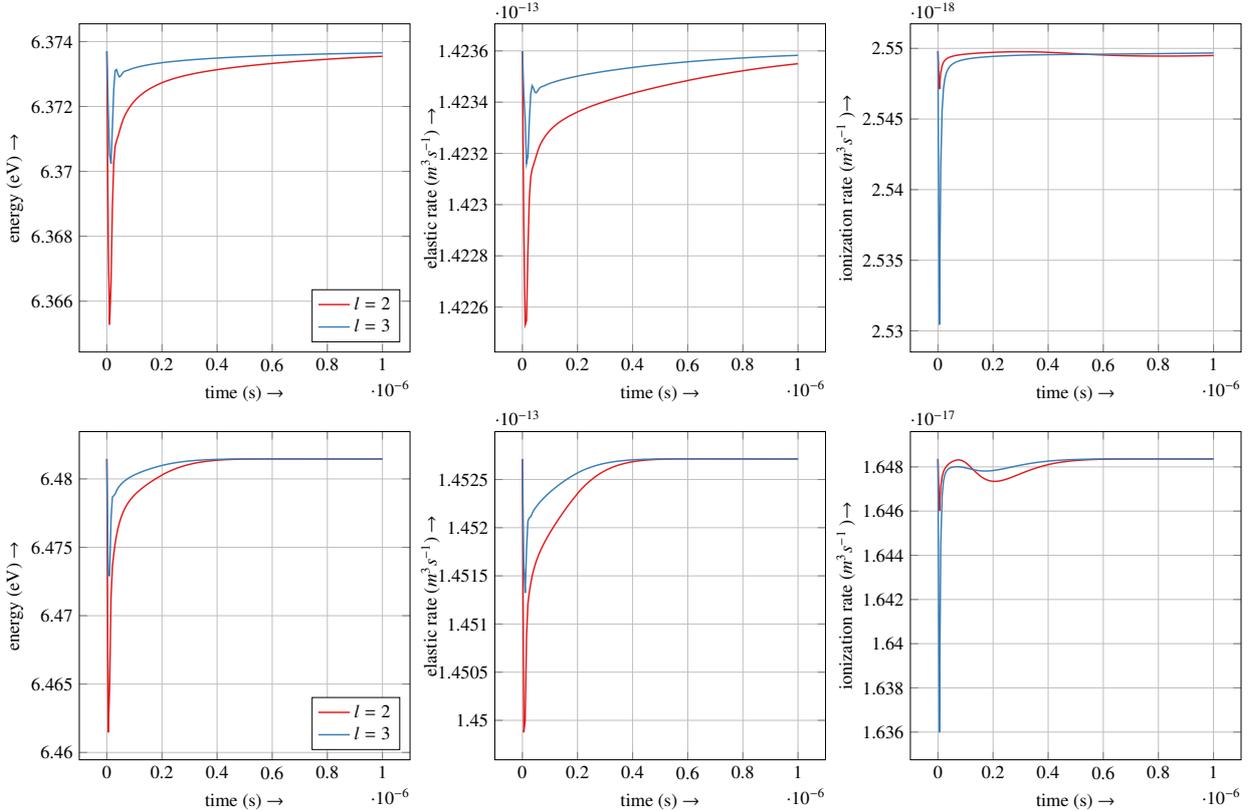

In this section, we analyze the relaxation time scales of the system if the higher $l$-modes are perturbed. The above study can be useful in understanding the importance of higher-order modes in more generic transport settings (i.e., spatially inhomogeneous BTE). For a given input parameters $E/n_0$ and $n_e/n_0$, we compute the steady-state solution and extract the $f_0^{steady}$ (i.e., $l$=0) component, and the initial distribution function for perturbing a specific $l$-mode is given by \Cref{eq:f_perturb}.
\begin{align}
	f^{(l)}(v,\vtheta, t=0) = 
	\begin{cases}
		f_0^{steady}(v) \text{ for } l=0 ,\\
		f_0^{steady}(v) + f_l(v) Y_{l0}(\vtheta) \text{ where } f_l(v) = f_0^{steady}(v)	\text{ for } l>0.	
	\end{cases} \label{eq:f_perturb}
\end{align}
The above initial condition evolves over time until the introduced perturbation dies out and the steady-state solution is reached. The observed relaxation time scales are inversely proportional to the electron temperature. Higher electron temperature results in faster energy transfer between particles, leading to shorter relaxation time scales. \Cref{fig:perturb_analysis} shows the relaxation time scales for $l=2$ and $l=3$ modes under specific input parameters. Even though the result in perturbations in the QoIs is relatively small, higher-order mode perturbations take significant time to settle down, especially under low temperatures. The above can be crucial in dealing with spatially inhomogeneous BTE, where any perturbations to higher-order modes get mixed spatially through spatial advection. In \Cref{fig:perturb_analysis} plotted QoIs explicitly depend on the $f_0$ and $f_1$ modes, while all the $l$ modes get coupled through the Boltzmann equation.

\subsection{Oscillatory electric field}
\begin{figure}[!tbhp]
\centering
  \resizebox{0.98\textwidth}{!}{
  \begin{tikzpicture}
				\begin{axis}[xlabel= time (s) $\rightarrow$, ylabel= energy (eV) $\rightarrow$, grid=major,width = 0.48\textwidth, height=0.48\textwidth, legend pos=north east]
					\addplot[-,a1  ,thick] table[x={time} , y ={energy}]{pde_runs/rs_nr128_lmax1_E3.22E+02_id_0.00E+00_Tg0.00E+00_dt2.00E-09.csv};
					\legend{transient}
				\end{axis}
			\end{tikzpicture}
			\begin{tikzpicture}
				\begin{axis}[xlabel= time (s) $\rightarrow$, ylabel= elastic reaction rate ($m^3s^{-1}$) $\rightarrow$, grid=major,width = 0.48\textwidth, height=0.48\textwidth, legend pos=north east]
					\addplot[-,a1  ,thick] table[x={time}, y ={g0}]    {pde_runs/rs_nr128_lmax1_E3.22E+02_id_0.00E+00_Tg0.00E+00_dt2.00E-09.csv};
					\addplot[-,a2  ,thick] table[x={time}, y ={g0_inp}]{pde_runs/rs_nr128_lmax1_E3.22E+02_id_0.00E+00_Tg0.00E+00_dt2.00E-09.csv};
					\legend{transient , interpolated}
				\end{axis}
			\end{tikzpicture}
			\begin{tikzpicture}
				\begin{semilogyaxis}[xlabel= time (s) $\rightarrow$, ylabel= ionization reaction rate ($m^3s^{-1}$) $\rightarrow$, grid=major,width = 0.48\textwidth, height=0.48\textwidth]
				\addplot[-,a1  ,thick] table[x={time}, y ={g2} ]	  {pde_runs/rs_nr128_lmax1_E3.22E+02_id_0.00E+00_Tg0.00E+00_dt2.00E-09.csv};
				\addplot[-,a2  ,thick] table[x={time}, y ={g2_inp} ]{pde_runs/rs_nr128_lmax1_E3.22E+02_id_0.00E+00_Tg0.00E+00_dt2.00E-09.csv};
				\end{semilogyaxis}
    \end{tikzpicture}
  }
  \caption{A simple illustration of the difference between the QoIs evolution with the transient solver compared against the QoIs reconstruction with interpolation using tabulated steady-state QoIs. We use the exact electron temperature evolution to compute the interpolated QoIs. We observe that the ionization rate coefficient is not well approximated  by the table-based interpolation.  \label{fig:transient_vs_steady_state}}
\end{figure}
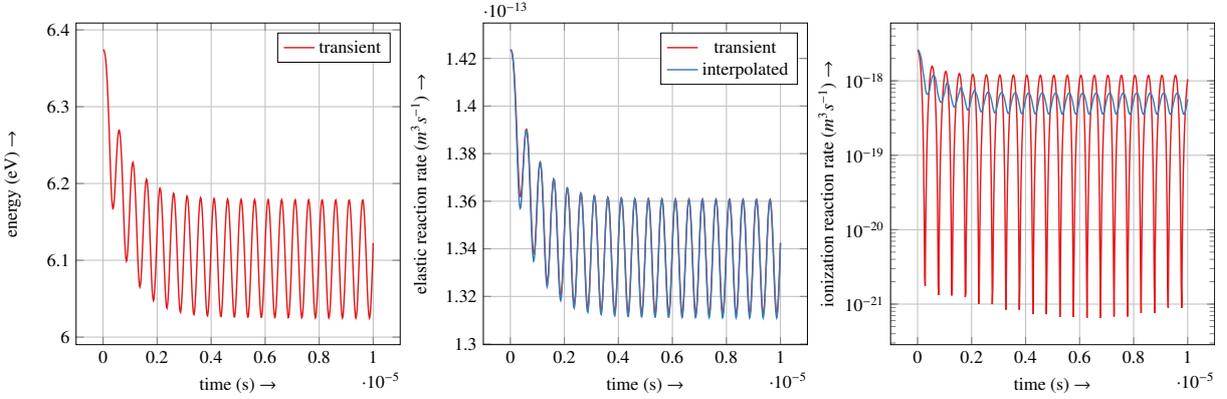

Typical LTP simulations build temperature-based lookup tables for electron kinetic coefficients. That is, we can compute steady-state EDF solutions for spatially homogeneous BTE with varying $E/n_0$ values, compute the resulting electron temperature and rate coefficients and the tabulate rate coefficients, mobility, and diffusivity as a function of electron temperature. These lookup tables can eliminate the direct BTE coupling. Here we explore the accuracy of the approach described above. Specifically, we explore the accuracy of temperature-based lookup tables for predicting electron kinetics under oscillatory electric fields without solving the BTE. In order to isolate errors due to lookup table interpolation, we use the exact electron temperature profile computed by solving BTE. \Cref{fig:transient_vs_steady_state} shows the discrepancies between the time-evolved QoIs (i.e., QoIs computed based on time-evolved BTE) compared against the interpolated QoIs assuming the exact evolution of the electron temperature profile (i.e., using transient electron temperature for lookup tables). With the lookup table approach, we observe wile  the elastic collision rate coefficient is well approximated  the ionization rate coefficient is not. This discrepancy can be explained by the sensitivity differences in elastic and ionization rate coefficients for the EDF tails. It follows that direct coupling of the BTE solver to the remaining flow is important in transient simulations. 

\section{Conclusions} \label{sec:conclusion}
We present a fast solver for spatially homogeneous electron Boltzmann equation. Our solver is based on the Galerkin discretization of the BTE. The presented framework supports electron-heavy binary collisions (i.e., elastic, excitation, ionization ) and electron-electron Coulomb interactions under constant and time-varying electric fields. The developed solver enables multi-term EDF approximations for steady-state and transient BTE solutions. Our methodology extends the state-of-the-art traditional two-term EDF approximation. We present numerical results showing the convergence of the developed Eulerian solver and verification results with the state-of-the-art \bolsig~code and in-house particle-in-cell DSMC code. Due to the limitations of the \bolsig~two-term EDF approximations, the DSMC code is used for the verification of the higher-order EDF correction modes. Additionally, we show that the relaxation time scales for the perturbed higher-order EDF modes are significant, especially in low-temperature regimes. Therefore, these higher-order correction modes may have a significant effect on spatially coupled BTE. In many LTP simulations, BTE-computed tabulated electron kinetics are used as a closure model for electron kinetics. To evaluate the above, we present modeling errors for table-based interpolation of electron kinetics under time-varying electric fields. Based on the above study, we can conclude that table-based interpolation is not ideal for electron kinetics that are heavily influenced by the EDF tails (i.e., specific rate coefficients such as ionization). The developed solver is implemented in Python and openly available at \href{https://github.com/ut-padas/boltzmann}{https://github.com/ut-padas/boltzmann}.



\appendix

\section{Eulerian approach}
\label{sec:deterministic_approach}
Here we summarize detailed derivations related to the Eulerian BTE solver. 
\subsection{Advection operator}
\label{subsec:advection}
In spherical coordinates we can write the gradient operator as \Cref{eq:adv_sph}. The electric field aligned on the z-axis is given by \Cref{eq:e_field}. 
\begin{align}
\nabla_{\vect{v}} 
&= \vrunit \frac{\partial}{\partial \vr}
+ \vthetaunit \frac{1}{\vr} \frac{\partial}{\partial \vtheta}
+ \vphiunit \frac{1}{\vr \sin\of{\vtheta}} \frac{\partial}{\partial \vphi} \label{eq:adv_sph}\\
\vect{E} &= E \hat{\vect{z}}  = E \left( \cos\of{\vtheta} \vrunit - \sin\of{\vtheta} \vthetaunit \right) \label{eq:e_field}
\end{align} With the above assumptions the advection term can be written as \Cref{eq:adv_form}.
\begin{align}
    \vect{E}\cdot \nabla_{\vect{v}} f = E \left( \cos\of{\vtheta} \frac{\partial f}{\partial \vr} - \sin\of{\vtheta} \frac{1}{\vr} \frac{\partial f}{\partial \vtheta} \right) \label{eq:adv_form}
\end{align} We use the recurrence relations given by \Cref{eq:ll_r1} and \Cref{eq:ll_r2} for the associated Legendre polynomials. 
\begin{align}
x P^m_l\of{x}   &=  \alpha_M\of{l,m} P^m_{l-1}\of{x} +  \beta_M\of{l,m} P^m_{l+1}\of{x}        \label{eq:ll_r1}       \\
(1-x^2) \frac{d}{dx} P^m_l\of{x} &= \alpha_D\of{l,m} P^m_{l-1}\of{x} - \beta_D\of{l,m} P^m_{l+1}\of{x}  \text{ where } \label{eq:ll_r2}\\
\alpha_M\of{l,m} = \frac{l+m}{2l+1} \text{ , } \beta_M\of{l,m} = \frac{l-m+1}{2l+1} \text{ , }&
\alpha_D\of{l,m} = \frac{(l+1)(l+m)}{2l+1} \text{ and } \beta_D\of{l,m} = \frac{l(l-m+1)}{2l+1}
\end{align} Using \Cref{eq:ll_r1} and \Cref{eq:ll_r2} and taking $x=\cos(\vtheta)$ we can show \Cref{eq:adv_rr_1} and \Cref{eq:adv_rr_2} respectively.
\begin{align}
\cos\of{\vtheta} Y_{lm}\of{\vtheta,\vphi}
&=
\cos\of{\vtheta} U_{lm} P^{|m|}_l\of{\cos\of{\vtheta}} \alpha_m\of{\vphi} \nonumber
\\
&=
U_{lm} \left( \alpha_M\of{l,|m|} P^{|m|}_{l-1}\of{\cos\of{\vtheta}} 
+ \beta_M\of{l,|m|} P^{|m|}_{l-1}\of{\cos\of{\vtheta}} \right) \alpha_m\of{\vphi} \nonumber
\\
&=
\underbrace{\alpha_M\of{l,|m|} \frac{U_{lm}}{U_{(l-1)m}}}_{A_M\of{l,m}} Y_{(l-1)m}\of{\vtheta,\vphi} 
+ \underbrace{\beta_M\of{l,|m|} \frac{U_{lm}}{U_{(l+1)m}}}_{B_M\of{l,m}} Y_{(l+1)m}\of{\vtheta,\vphi} \nonumber\\
&=
A_M\of{l,m} Y_{(l-1)m}\of{\vtheta,\vphi} 
+ 
B_M\of{l,m} Y_{(l+1)m}\of{\vtheta,\vphi} \label{eq:adv_rr_1}
\end{align}
\begin{align}
-\sin\of{\vtheta} \frac{d}{d\vtheta} Y_{lm}\of{\vtheta,\vphi}
&=
-\sin\of{\vtheta} U_{lm} \frac{d}{d\vtheta} P^{|m|}_l\of{\cos\of{\vtheta}} \alpha_m\of{\vphi} 
\nonumber \\
&=
\left(\sin\of{\vtheta}\right)^2 U_{lm} \frac{d}{d\cos\of{\vtheta}} P^{|m|}_l\of{\cos\of{\vtheta}} \alpha_m\of{\vphi}
\nonumber \\
&=
\left(1 - \left(\cos\of{\vtheta}\right)^2\right) U_{lm} \frac{d}{d\cos\of{\vtheta}} P^{|m|}_l\of{\cos\of{\vtheta}} \alpha_m\of{\vphi}
\nonumber \\
&=
U_{lm} \left( \alpha_D\of{l,|m|} P^{|m|}_{l-1}\of{\cos\of{\vtheta}} 
+ \beta_D\of{l,|m|} P^{|m|}_{l-1}\of{\cos\of{\vtheta}} \right) \alpha_m\of{\vphi} \nonumber
\\
&=
\underbrace{\alpha_D\of{l,|m|} \frac{U_{lm}}{U_{(l-1)m}}}_{A_D\of{l,m}} Y_{(l-1)m}\of{\vtheta,\vphi} 
+ \underbrace{\beta_D\of{l,|m|} \frac{U_{lm}}{U_{(l+1)m}}}_{B_D\of{l,m}} Y_{(l+1)m}\of{\vtheta,\vphi} \nonumber
\\
&=
A_D\of{l,m} Y_{(l-1)m}\of{\vtheta,\vphi} 
+ 
B_D\of{l,m} Y_{(l+1)m}\of{\vtheta,\vphi} \label{eq:adv_rr_2}
\end{align} By substituting \Cref{eq:adv_rr_1} and \Cref{eq:adv_rr_2} to \Cref{eq:adv_form} we can write \Cref{eq:adv_term_aa}.

\begin{multline}
    \vect{E} \cdot \nabla_{\vect{v}} f = E 
    \sum_{klm} f_{klm} 
    \Bigg(
    \left( A_M\of{l,m} \frac{d}{d\vr}\phi_{k}\of{v} 
    + A_D\of{l,m} \frac{1}{\vr}\phi_{k}\of{v} \right)
    Y_{(l-1)m}\of{\vtheta,\vphi} 
    \\
    +
    \left( B_M\of{l,m} \frac{d}{d\vr}\phi_{k}\of{v} 
    + B_D\of{l,m} \frac{1}{\vr}\phi_{k}\of{v} \right)
    Y_{(l+1)m}\of{\vtheta,\vphi} 
    \Bigg) \label{eq:adv_term_aa}
\end{multline} By projecting the above on to the test basis and using the orthogonality of the spherical basis we can write \Cref{eq:adv_ws_a}.
\begin{multline}
    \myint_{R} \myint_{S^2} \Big(\vect{E}\cdot \nabla_{\vect{v}} f \Big) \phi_{p}\of{v} Y_{qs}\of{\theta,\phi} 
    \vr^2  \diff{\omega_v} \diff{\vr}\\
    = E 
    \myint 
    \sum_{k}
    \Bigg(
    f_{k(q+1)s} 
    \left( A_M\of{q+1,s} \frac{d}{d\vr}\phi_{k}\of{v} 
    + A_D\of{q+1,s} \frac{1}{\vr}\phi_{k}\of{v} \right)
    \\
    + f_{k(q-1)s} \left( B_M\of{q-1,s} \frac{d}{d\vr}\phi_{k}\of{v} 
    + B_D\of{q-1,s} \frac{1}{\vr}\phi_{k}\of{v} \right)
    \Bigg)
    \phi_{p}\of{v}
    \vr^2 \diff{\vr} \label{eq:adv_ws_a}
\end{multline} 

\subsection{Electron-heavy collision operator simplifications}
\label{subsec:collop_simplifications}
Let $\vect{v} = (v, v_\theta, v_\phi) $ and scattering solid angle $\vect{\omega} = (\chi, \phi)$. Further, let $\Phi_{klm}\of{\vect{v}} = \phi_{k}\of{v} Y_{lm}\of{\vtheta,\vphi}$ and $\Psi_{pqs}\of{\vect{v}} = \phi_{p}\of{v} Y_{qs}\of{\vtheta,\vphi}$ be the trial and test basis functions. Let $\vect{v}^{\prime} = (v^\prime, v_\theta^\prime, v_\phi^\prime) =\vect{v}^{post}\of{\vect{v},\vect{0},\vect{\omega}}$ be the post scattering velocity. We can show the angular coordinate of the post scattering velocity obeys \Cref{eq:scattering_post}.  
\begin{align}
\cos v_\theta^\prime = \cos v_\theta \cos\chi + \sin v_\theta \sin \chi \cos\of{v_\phi-\phi} \label{eq:scattering_post}
\end{align} Using the spherical harmonic addition theorem we can write \Cref{eq:sph_addition_thm} 
\begin{align}
P_l^{0} = P_l\of{\cos v_\theta^\prime} =  P_l\of{\cos v_\theta} P_l\of{\cos v_\phi} + 2\sum_{m=1}^{l} P_l^{m}\of{\cos v_\theta} P_l^{m}\of{\cos v_\phi} \cos \of{ m (v_\phi-\phi)} \label{eq:sph_addition_thm}
\end{align} The discretized weak form of the collision operator approximation can be written as \Cref{eq:c_en_mat_a}.
\begin{equation}
    {[C_{en}]}^{pqs}_{klm} = n_0 \myint_{R} \myint_{S^2} \myint_{S^2} 
    v^2 \sigma\of{\vect{v},\vect{\omega}} \phi_{k}\of{v} Y_{lm}\of{\vtheta,\vphi}
    \of{\phi_{p}\of{u} Y_{qs}\of{\utheta,\uphi} -\phi_{p}\of{v} Y_{qs}\of{\vtheta,\vphi}} \diff{\omega_v} \diff{\omega}\diff{v} \label{eq:c_en_mat_a}
\end{equation} Assuming isotropic scattering and writing $\sigma\of{v,\vect{\omega}}=\frac{\sigma_T\of{v}}{4\pi}$  we can further simplify and get \eqref{eq:c_en_simp_1}.
\begin{align}
[C_{en}]^{pqs}_{klm} &= \frac{n_0}{4\pi} \myint_{R} v^2 \sigma_T\of{v} \phi_k\of{v} 
\myint_{S^2} \myint_{S^2} Y_{lm}\of{v_\theta,v_\phi}
\of{\phi_{p}\of{v^\prime} Y_{qs}\of{v^\prime_\theta, v^\prime_\phi} - \phi_{p}\of{v} Y_{qs}\of{v_\theta, v_\phi}} \diff{\omega} \diff{\omega_{v}}\diff{v} \label{eq:c_en_simp_1}
\end{align} Let us write the angular integral part separately as \Cref{eq:c_en_angular}. 
\begin{align}
W^{qs}_{lm} \of{v}= \myint_{S^2} \myint_{S^2}  Y_{lm}\of{v_\theta,v_\phi}
\of{\phi_{p}\of{v^\prime} Y_{qs}\of{v^\prime_\theta, v^\prime_\phi} - \phi_{p}\of{v} Y_{qs}\of{v_\theta, v_\phi}} \diff{\omega} \diff{\omega_{v}} = {W^{qs}_{lm}}^{+} - {W^{qs}_{lm}}^{-} \label{eq:c_en_angular}
\end{align}Due to the orthogonality of the spherical harmonics, we can simplify ${W^{qs}_{lm}}^{-}$ as in \Cref{eq:Wm}. 
\begin{align}
{W^{qs}_{lm}}^- =  \phi_{p}\of{v} \myint_{S^2} \myint_{S^2}  Y_{lm}\of{v_\theta,v_\phi} Y_{qs}\of{v_\theta, v_\phi}  \diff{\omega} \diff{\omega_{v}} 
= \phi_p\of{v} 4\pi \delta_{ql} \delta_{sm} \label{eq:Wm}
\end{align} Assuming the azimuthal symmetry, we can simplify the term ${W^{qs}_{lm}}^{+}$ with application of spherical harmonics addition theorem and derive \eqref{eq:Wp}.  
\begin{align}
{W^{q0}_{l0}}^+ &= \myint_{S^2} \myint_{S^2}  Y_{lm}\of{v_\theta,v_\phi} \phi_{p}\of{v^\prime} Y_{qs}\of{v^\prime_\theta, v^\prime_\phi} \diff{\omega} \diff{\omega_{v}} \nonumber \\
& = \phi_{p}\of{v^\prime} U_{l0} U_{q0} \myint_{S^2} \myint_{S^2}  P_l\of{\cos v_\theta} P_q\of{\cos v^\prime_\theta} \diff{\omega} \diff{\omega_{v}} \nonumber\\
& = \phi_{p}\of{v^\prime} U_{l0} U_{q0} \myint_{S^2} \myint_{S^2}  P_l\of{\cos v_\theta} \of{P_q\of{\cos v_\theta} P_q\of{\cos \chi} + \cdots} \diff{\omega} \diff{\omega_{v}} \nonumber\\
& = \phi_{p}\of{v^\prime} U_{l0} U_{q0} \myint_{S^2} P_l\of{\cos v_\theta} P_q\of{\cos v_\theta} \diff{\omega_{v}} \myint_{S^2} P_0\of{\cos \chi}P_q\of{\cos \chi} \diff{\omega}\nonumber\\
& = 4\pi \phi_{p}\of{v^\prime} \sqrt{\frac{2l+1}{2q+1}} \delta_{ql} \delta_{q0} \label{eq:Wp}
\end{align}Therefore, we can write the simplified electron-heavy collision operator as \Cref{eq:en_simplified}.
\begin{align}
[C_{en}]^{pq0}_{kl0}  = n_0 \myint_{R} v^2 \sigma_T(v) \phi_k\of{v} \delta_{ql} \of{\phi_p\of{v^\prime}\delta_{q0} - \phi_p\of{v}} \diff{v} \label{eq:en_simplified}
\end{align}

\subsection{Coulomb collisions}
\label{subsec:ee_collisions}
For coordinate system given by $x_1 = v, x_2 = \cos{\vtheta}, x_3 = \vphi$ we can write the line element by \Cref{eq:sph_line_element}. 
\begin{align}
    ds^2   = dv^2 + v^2(1-\mu^2)^{-1} d\mu^2 + v^2(1-\mu^2) d\phi^2 \label{eq:sph_line_element}
\end{align}. Using the Einstein summation notation (i.e., repeated matching up and down indices are summed over) we can write the divergence and the Hessian operator acting on scalar as \Cref{eq:div} and \Cref{eq:hessian} respectively where $D_i$ denotes the covarient derivative, $\Gamma^{k}_{ij}$ denotes the Christoffel symbols, $a=\det{a_{ij}}$ (i.e., determinant of the metric)  and $D_i D_j$ denotes the Hessian operator. 
\begin{align}
    D_i h^i &= \frac{1}{\sqrt{a}} \partial_i \of{\sqrt{a} h^i} \label{eq:div} \\
    D_i D_j g &= \partial^2_{ij} g - \Gamma^k_{ij} \partial_k g \label{eq:hessian}
\end{align} 
\section{DSMC Approach}
\subsection{Calculating $l$-modes}

We use spherical harmonics to approximate the EDF $f(v_r, v_\theta)$ as in \Cref{eq:sph_expansion}.
\begin{equation}
    f(v_r, v_\theta) = \sum_{l=0}^{l'} f_l(v_r) Y_{l0}(v_\theta). \label{eq:sph_expansion}
\end{equation}
Contrast to the Eulerian approach DSMC approach uses particles to represent EDF. To compute the $f_l$ modes in DSMC approach we project the DSMC solution to spherical harmonics, where integral in \Cref{eq:sph_integrals} are computed using Monte Carlo approximation. 
\begin{equation}
f_l(v_r) = 2 \pi \int_0^{2\pi} f(v_r, v_{\theta}) Y_{l0} \sin(v_{\theta}) d\theta \label{eq:sph_integrals}.
\end{equation}









\bibliographystyle{elsarticle-num}
\bibliography{bte}







\end{document}